\documentclass[a4paper,fleqn,usenatbib]{mnras}
\usepackage{graphicx}
\usepackage[dvipsnames]{xcolor}
\usepackage{siunitx}
\usepackage{verbatim}
\usepackage[english]{babel}
\usepackage[T1]{fontenc}
\usepackage{ae,aecompl}

\newcommand{\bll}{$L_{\textrm{\scriptsize BLL}}$}
\newcommand{\fsrq}{$L_{\textrm{\scriptsize FSRQ}}$}

\newcommand{\bann}{BLL$_{\textrm{\scriptsize ANN}}$}
\newcommand{\fann}{FSRQ$_{\textrm{\scriptsize ANN}}$}
\newcommand{\bfgl}{BLL$_{\textrm{\scriptsize 3FGL}}$}
\newcommand{\ffgl}{FSRQ$_{\textrm{\scriptsize 3FGL}}$}
\newcommand{\bcfgl}{BCU$_{\textrm{\scriptsize 3FGL}}$}
\newcommand{\btann}{BLL$_{\textrm{\scriptsize 3FGL+ANN}}$}
\newcommand{\ftann}{FSRQ$_{\textrm{\scriptsize 3FGL+ANN}}$}
\newcommand{\bctann}{BCU$_{\textrm{\scriptsize 3FGL+ANN}}$}
\newcommand{\bcann}{BCU$_{\textrm{\scriptsize ANN}}$}
\newcommand{\bflapann}{\textit{B-FlaP$_{\textrm{\scriptsize ANN}}$}}

\usepackage{textcomp}
\DeclareTextSymbolDefault{\degreesymbol}{TS1} 
\DeclareTextSymbol{\degreesymbol}{TS1}{176} 
\DeclareRobustCommand{\textdegree}{\ifmmode\mbox{\degreesymbol}\else\degreesymbol\fi}

\usepackage[switch]{lineno}

\title[Blazar Flaring Patterns (B-FlaP)]{ Blazar Flaring Patterns (B-FlaP)\\ Classifying  Blazar Candidate  of Uncertain type  
in the Third Fermi-LAT catalog by Artificial Neural Networks}

\author[G. Chiaro et al.]{
G. Chiaro$^{1}$\thanks{E-mail: chiaro@pd.infn.it},
D. Salvetti$^{2}$\thanks{E-mail: salvetti@iasf-milano.inaf.it},
G. La Mura$^{1}$,
M. Giroletti$^{3}$,
D. J. Thompson$^{4}$
\newauthor
and D. Bastieri$^{1}$
\\
$^{1}$Dip. Fisica \& Astronomia G. Galilei - Universit\`a di Padova, I-35131, Padova, Italy\\
$^{2}$INAF -Istituto di Astrofisica Spaziale e Fisica Cosmica , I-20133, Milano, Italy\\
$^{3}$INAF-Institute of Radioastronomy, I-40129, Bologna, Italy\\ 
$^{4}$NASA Goddard Space Flight Center , Greenbelt, MD,  USA
}

\date{Accepted 2016 July 21. Received 2016 July 21; in original form 2016 May 12}

\pubyear{2016}

\begin{document}
\label{firstpage}
\pagerange{\pageref{firstpage}--\pageref{lastpage}}
\maketitle

\begin{abstract}  
The {\it Fermi} Large Area Telescope (LAT) is currently the most important facility for investigating the GeV $\gamma$-ray sky.  With {\it Fermi} LAT more than three thousand  $\gamma$-ray sources have been discovered so far. 1144 ($\sim40\%$) of the  sources are active galaxies of the blazar class, and 573 ($\sim20\%$) are listed as Blazar Candidate of Uncertain type  (BCU), or sources without a conclusive classification. We use the Empirical Cumulative Distribution Functions (ECDF) and the Artificial Neural Networks (ANN) for a fast method of screening and classification for BCUs based on data collected at $\gamma$-ray energies only, when rigorous multiwavelength analysis is not available. Based on our method, we classify 342 BCUs as BL Lacs and 154 as FSRQs, while 77 objects remain uncertain. Moreover, radio analysis and direct observations in ground-based optical observatories are used as counterparts to the statistical classifications to validate the method. 
This approach is of interest because of the increasing number of unclassified sources in {\it Fermi} catalogs and  because blazars and in particular their subclass High Synchrotron Peak (HSP) objects are the main targets of  atmospheric Cherenkov telescopes. 
\end{abstract}


\section{Introduction} 

 Blazars are active galactic nuclei (AGN) with a radio-loud behavior and a relativistic jet pointing toward the observer. \citep{Abdo01} 
These sources are divided into two main classes: BL Lacertae objects (BL Lacs) and Flat Spectrum Radio Quasars (FSRQs), which show very different optical spectra even if in other wavebands they are similar.
FSRQs have strong, broad emission lines at optical wavelengths, while BL Lacs show at most weak emission lines, sometimes display absorption features, and can also be completely featureless. 
 Compact radio cores, flat radio spectra, high brightness temperatures, superluminal motion, high polarization, and strong and rapid variability are commonly found in both BL Lacs and FSRQs.  
Blazars emit variable, non-thermal radiation across the whole electromagnetic spectrum, which includes two components forming two broad humps in a $\nu f{_\nu}$ representation. The low-energy one is attributed to synchrotron radiation, and the high-energy one is usually thought to be due to inverse Compton radiation.  See \citet{Ghisellini} for a recent review of the properties of $\gamma$-ray AGN. 
Blazars can also be classified into different subclasses based on the position of the peak of the synchrotron bump in their spectral energy distribution (SED), namely, low frequency peaked (LSP or sources with $\nu^{S}_{peak}$ < $10^{14}$ Hz), 
intermediate frequency peaked (ISP or sources with $10^{14}$ Hz < $\nu^{S}_{peak}$ < $10^{15}$ Hz)
and high frequency peaked (HSP or sources with $\nu^{S}_{peak}$ > $10^{15}$ Hz ) \citep{Abdo02}. 
This subclassification suggests the possibility that the $\gamma$-ray properties of the sources may lead to constraints on the type of objects responsible for the radiation especially in view of the increasing number of detections obtained by the {\it Fermi}  Large Area Telescope (LAT) that still have to be properly classified.

The Third {\it Fermi}-LAT Source Catalog (3FGL)\citep{ace15} listed  3033  $\gamma$-ray sources collected in four years of operation, from 2008 August 4 (MJD 54682) to 2012 July 31 (MJD 56139). 3FGL covers the full sky.  1144 sources are identified or associated with galaxies of the blazar class. 660 are BL Lacs and 484 are FSRQs. 3FGL includes also 573 Blazar Candidates of Uncertain type (BCUs). 
Because of the difficulty of having extensive optical observation campaigns for full classification of blazars, if we compare the 3FGL with previous catalogs released by the LAT collaboration we can see a significant increase of the number of unclassified blazars. In Table 1 we show the growth of the number of blazar-class sources  in {\it Fermi}-LAT  catalogs and the relative fraction of each blazar source subclass. The percentage  of BCUs within the blazar sample increased from 13.8$\%$ in 1FGL  to 33.4$\%$ in 3FGL. Although the detailed multiwavelength analysis necessary for unambiguous classification has been done and is continuing for many of these \citep{alvarez}, a first classifying screening of BCUs, as our method proposes, can be very useful for the blazar scientific community.

\begin{table}
\caption{Blazar-class source distribution in {\it Fermi}-LAT catalogs and the relative fraction of each blazar source subclass}
\begin{center}
\begin{footnotesize}
\begin{tabular}{lccr}
\hline
\hline
\bf{Class} &\bf{1FGL} & \bf{2FGL} & \bf{3FGL} \\
\hline
BLL & 295 (44.4\%) & 436 (41\%) & 660 (38.4\%)\\ 
FSRQ & 278 (41.8\%) & 370 (34.8\%) & 484 (28.2\%)\\
BCU & 92 (13.8\%) & 257 (24.2\%) & 573 (33.4\%)\\
\hline
Total & 665 & 1063 & 1717\\ 
\hline
\end{tabular}
\end{footnotesize}
\end{center}
\end{table}

The aim of this work is to find a simple estimator in order to classify BCUs and, when it is possible, to identify high-confidence HSP candidates. The present generation of Imaging Atmospheric Cherenkov Telescopes (IACTs), such as VERITAS, H.E.S.S. and MAGIC, has opened  the realm of ground-based $\gamma$-ray astronomy in the Very High Energy range (VHE: E $> $100 GeV). The Cherenkov Telescope Array (CTA) will explore our Universe in depth in this energy band and lower. For a recent review of present and future Cherenkov telescopes, see \citep{deNaurois}. The BL Lac HSP sources are the most numerous class of TeV sources. The TeV catalog \citep{horan} reports 176 TeV sources. 46 of them are HSP BL Lacs  and only 5 FSRQs, therefore the ability to correctly identify HSP objects will be very important for the Cherenkov scientific community and in the determination of CTA targets, in order to increase the rate of detections, since IACTs have a small field of view.The novelty of the present approach is that our study  relies exclusively on variability data collected at  $\gamma$-ray energies where {\it Fermi}-LAT is most sensitive (0.1 -- 100 GeV) and it remains totally independent from other data at different wavelengths.
The paper is laid out as follows: in Section \ref{sec:LAT}, we present the $\gamma$-ray data and the ECDF light curves considered for our analysis; in Sect.~\ref{sec:ANN}, we describe the use of artificial neural networks, and in Sect.~\ref{sec:results} we present the results of the ANN analysis. In Sect.~\ref{sec:class} we present a summary of the results of our classification of BCUs listed in the 3FGL {\it Fermi}-LAT  and we highlight the most promising HSP candidates. In Sect.~\ref{sec:multiwave} we test our method comparing the predicted classifications with additional data, obtained through optical spectroscopy and radio observations. We summarize our conclusions in Sect.~\ref{sec:conclusions}.

\section {Gamma-ray Data} 
\label{sec:LAT}
 
\subsection{The Large Area Telescope}

The LAT is the primary instrument on the {\it Fermi Gamma-ray Space Telescope}, launched by NASA on 2008 June 11 and it is the first imaging {\bf GeV $\gamma$-ray observatory} able to survey the entire sky every day at high sensitivity orbiting the Earth every 96 minutes. The \emph{Fermi} LAT is a pair-conversion telescope with a precision converter-tracker and calorimeter. It measures the tracks of the electron and positron that result when an incident $\gamma$ ray undergoes pair-conversion and measures the energy of the subsequent electromagnetic shower that develops in the telescope's calorimeter \citep{FLAT}. Data obtained with \emph{Fermi}-LAT permit rapid notification and facilitate monitoring of variable sources such as the BCUs that we consider in this study. In this paper we used the monthly $\gamma$ flux value from LAT 4-year Point Source Catalog (3FGL) and the Fermi Science Support Center (FSSC) for any other data\footnote{\tt http://fermi.gsfc.nasa.gov/ssc/data/access/lat/4yr\_catalog/}.

\subsection{B-FlaP: Blazar Flaring Patterns } 
\begin{figure*}
\begin{center}
{\includegraphics[width=0.8\textwidth]{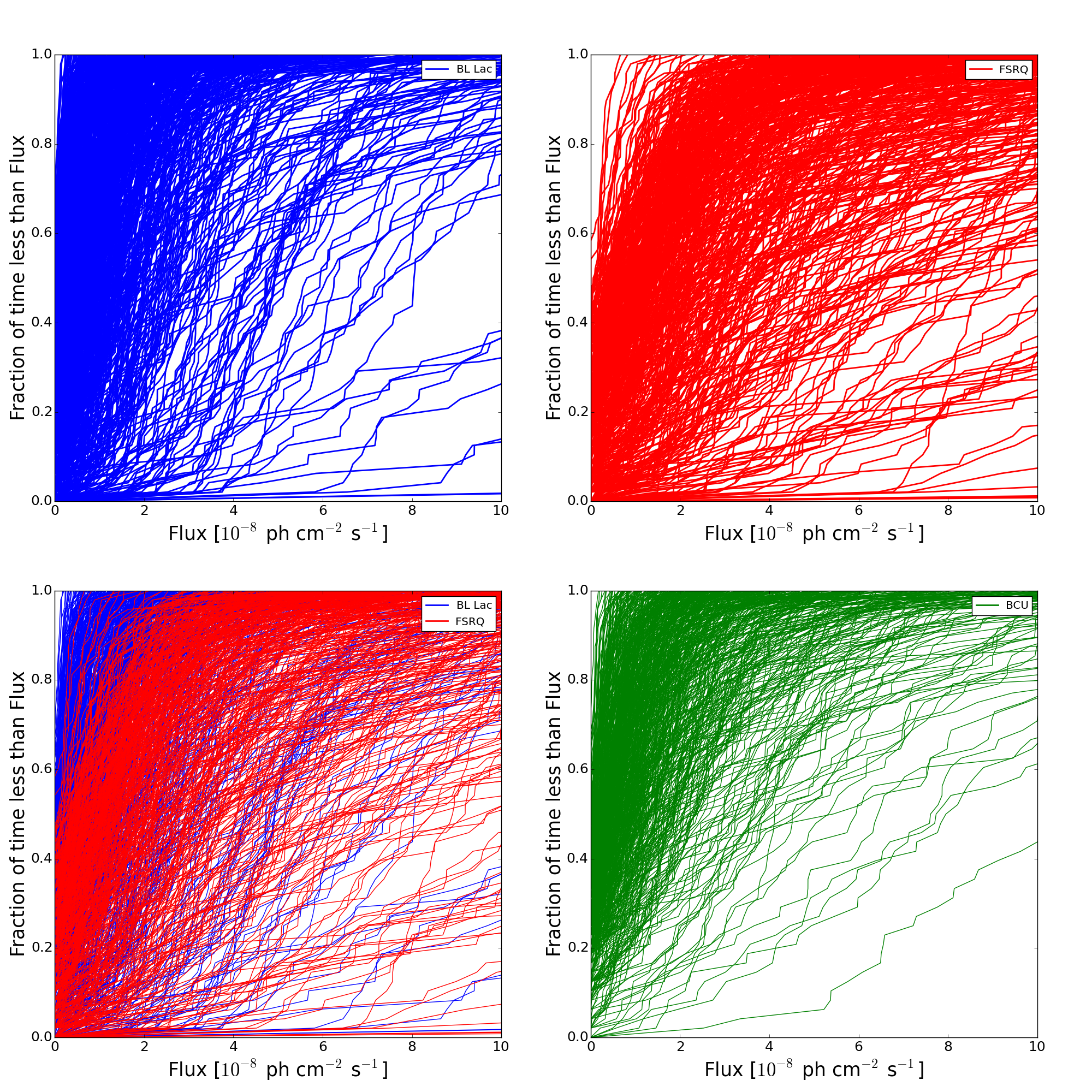}} 
\caption{ECDF plots of {\it Fermi} Blazars: BL Lacs (top left), FSRQs (top right) , BL Lac and FSRQ overlap (bottom left), BCUs (bottom right). 
The cumulative percentage of bins with flux below a given level is shown as a function of the 0.1 -- 100 GeV flux in a bin, in units of 10$^{-8}$ ph cm$^{-2}$  s$^{-1}$.\label{<Fig.1>}}
\end{center}
\end{figure*}

Variability is one of the defining characteristics of blazars \citep{Paggi01}. We considered the light curves of the blazar sources evaluated with monthly binning, as reported in 3FGL catalog, and with these data we designed the basic structure of the B-FlaP method.\\
The original idea was to compare the $\gamma$-ray light curve of the source under investigation with a \emph{template} classified blazar class light curve, then measure the difference in a proper metric.  
Typically $\gamma$-ray AGN are characterized by fast {\bf flaring} that could alter significantly the light curve and could make the comparison difficult.  In addition, different flux levels could hide the actual similarity of light curves. As first approach of this study we compute the Empirical Cumulative Distribution Function (ECDF) of the light curves \citep{KS}.
We  constructed the percentage of time when a source was below a given flux by sorting the data in ascending order of flux and then compared the ECDFs of BCUs with the ECDFs of blazars whose class is already established, (\S\ref{sec:LAT}). This is our variation of the Empirical Cumulative Distribution Function (ECDF) method. 
In Fig.~\ref{<Fig.1>} we show the ECDF plots for 3FGL blazars and BCUs.  In principle, differences due to the flaring patterns of BL Lacs and FSRQ appear in two ways:  (1) the  flux where the percentage reaches 100 represents the brightest flare seen for the source; and (2) the shape of the cumulative distribution curve reveals information about the flaring pattern, whether the source had one large flare, multiple flares, or few flares.The BL Lacs have fewer large flares than the FSRQs, and the FSRQ curves are more jagged, suggesting multiple flares compared to the smoother BL Lac curves.  
The difference between the classes is observed when we plotted the two blazar classes together.  At the bottom left of Fig.~\ref{<Fig.1>} is shown the significant overlap between the types where it is hard to distinguish individual objects, and there are outliers that extend beyond the range of the plots, but it is  possible to recognize on the top left of the diagram a specific area where the overlap between BL Lac and FSRQ is minimal. This area, at values of the flux less than $\sim$ 2.5 $\times$ 10$^{-8}$ ph cm$^{-2}$  s$^{-1}$, could lead to a first \emph{qualitative} recognition of BL Lac  objects.
In B-FlaP, special attention is needed for upper limits, which arise whenever light curves are constructed with fixed binning, as is the case here. They can be naturally incorporated into the current ECDF method, as the points plotted in the diagrams are the percentage of time that the source is below a given flux value.  Nevertheless, upper limits could introduce biases, skewing the cumulative distribution toward higher  percentages. Upper limits could be avoided entirely by producing light curves with adaptive binning \citep{Benoit02}, 
a technique that could be implemented into a possible follow-up study. For this reason  and because the ECDF plots represent only a proof of concept of the whole method, we follow up the ECDF first analysis with an Artificial Neural Network analysis (ANN)  by an original algorithm developed to distinguish the single BCU object and to give its likelihood to be a BL Lac or a FSRQ. \\
The reasons for the flaring patterns differences between BL Lacs and FSRQ are very likely connected with the processes occurring at the base of the jet, where the largest concentration of relativistic particles and energetic seed photons are expected. While in FSRQs accretion onto the central black hole produces a prominent and variable spectrum, characterized by continuum and emission-line photons, usually accompanied by the ejection of relativistic blobs of plasma in the jet, BL Lacs do not show such kind of activity and most of the observed radiation originates within the jet itself. As a consequence, the production of $\gamma$-ray emission through inverse Compton (IC) scattering can change much more dramatically in FSRQs  than in BL Lac-type sources, where the contribution of the central engine to the seed radiation field is weaker \citep{VAR}

\subsubsection{High Synchrotron Peak blazar} 

With reference to the aim of this study we applied the same ECDF technique to the blazar subclasses. 
Using the Third Catalog of Active Galactic Nuclei detected by Fermi-LAT \citep[3LAC,][]{Ackermann2015}, we collected information about classification and SED distribution of the blazars. The third release of the catalog considers only 1591 AGN detected at |\emph{b}| >$\ang{10}$ where \emph{b} is the Galactic latitude, 289 are classified sources as HSP on the basis of their SED, where 286 of them are represented by BL Lac objects and 3 by FSRQs. 160 of the 573 BCUs are HSP suspects. For all the other data in this study we referred to 3FGL.
While ISP and LSP blazars show the most variable patterns and can belong to both the BL Lac or FSRQ families, HSP objects are characterized by nearly constant emission. 

In Fig.~2 we plotted the ECDF for 3LAC HSPs versus FSRQs. As we expected, because of the fact that HSPs are almost exclusively represented by BL Lac objects,  the HSPs went through the BLL clean area at the upper left corner of the plot.
Even if ISP and LSP contamination is not negligible (Fig.~3), the result observed in Fig.~2 suggests \emph{the potential ability} of ECDF B-FlaP to identify a flux range at the 100th percentile (less than $\sim$ 2.0 $\times$ 10$^{-8}$ ph cm$^{-2}$  s$^{-1}$) where it is possible to not only determine the blazar class but also to tentatively assign the HSP subclass for a BCU source.

However, even here, visual inspection of the curves in all the ECDF figures suggests that the shape of the curve does not show major differences between the observed blazar classes. In order to improve the analysis we used the same ANN algorithm developed for BCUs  for the HSP classification.

\begin{figure}
\resizebox{\hsize}{!}{\includegraphics{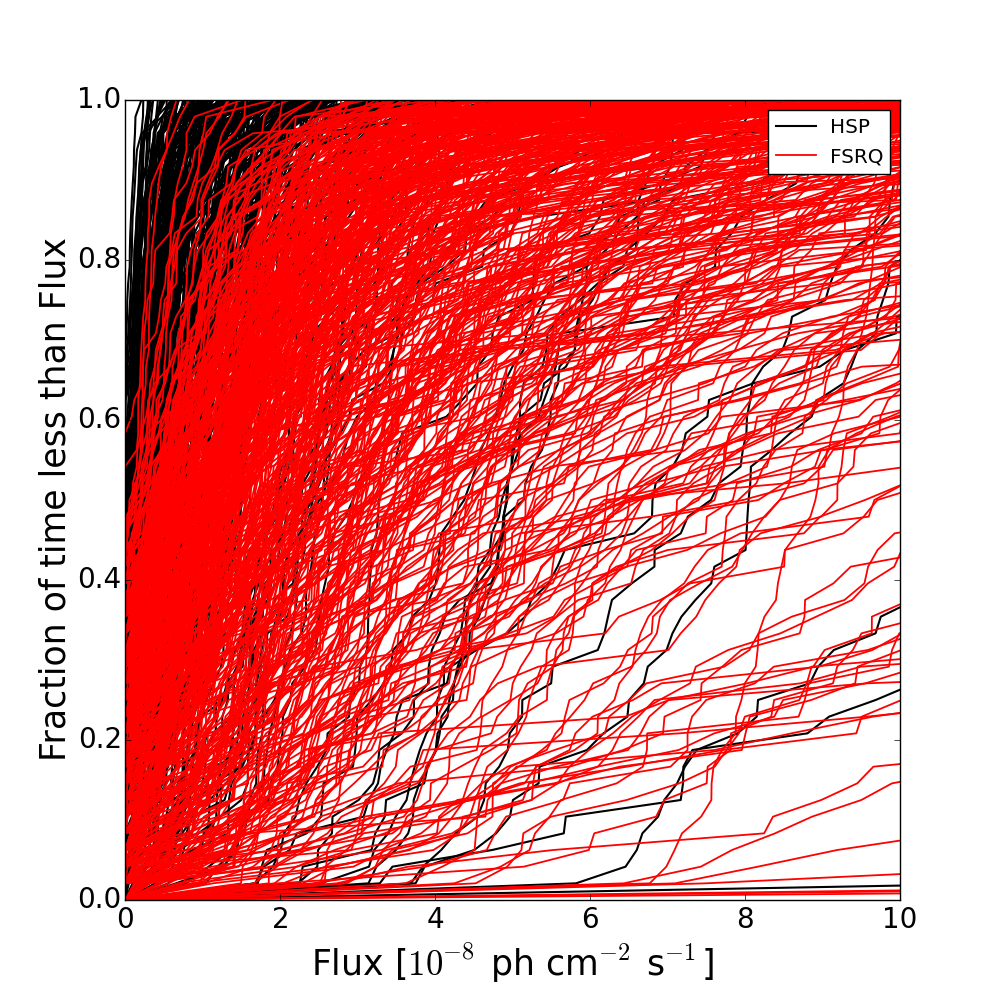}} 
\caption{ECDF for HSPs (black) and FSRQs (red), using the same construction and scale as Fig.1}
\end{figure}

\begin{figure}
\resizebox{\hsize}{!}{\includegraphics{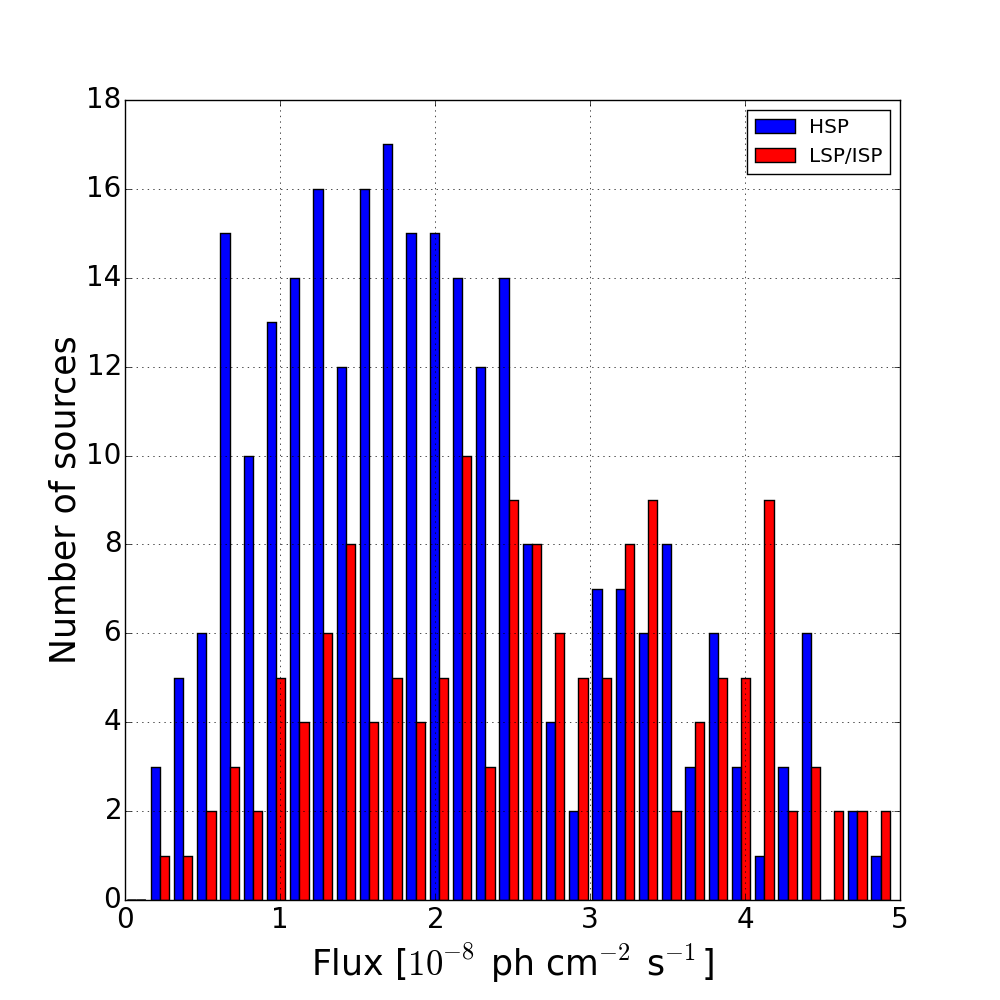}} 
\caption{LSPs and ISPs (red) versus HSPs (blue). Flux distribution at the 100th percentile. Even if, in the considered $\gamma$-ray flux range the number of HSPs is greater than the other subclasses, the degree of contamination cannot be overlooked.}
\end{figure}

\section{Artificial Neural Networks}
\label{sec:ANN}

In this section we describe the use of Artificial Neural Networks (ANNs) as a  promising method to classify blazar of uncertain types on the basis of their EDCF extracted from their $\gamma$-ray light curves.

\begin{figure}\resizebox{\hsize}{!}{\includegraphics{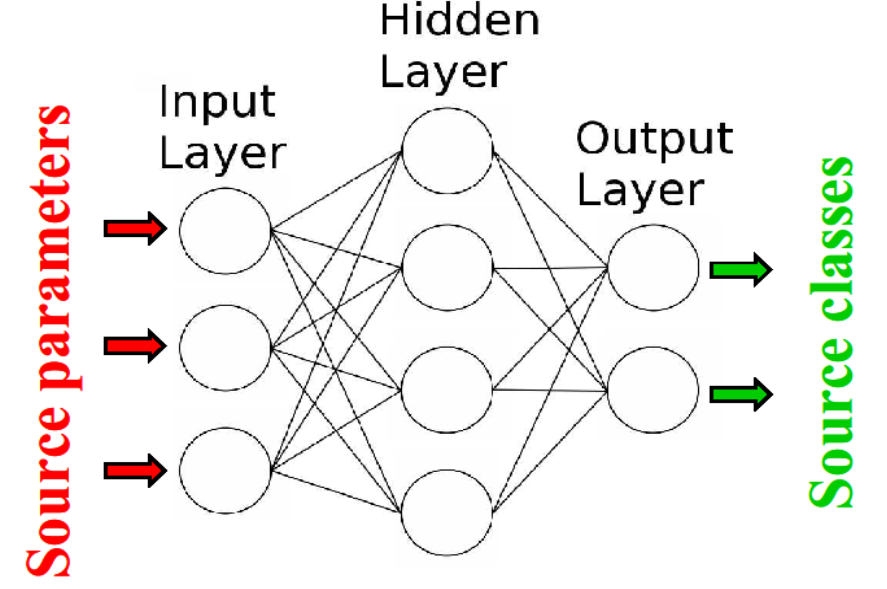}}
\caption{Schematic view of a Two Layer Perceptron (2LP), the Artificial Neural Network architecture we used for our analysis. Data enter the 2LP through the nodes in the input layer. The information travels from left to right across the links and is processed in the nodes through an activation function. Each node in the output layer returns the likelihood of a source to be a specific class.}\label{ann}
\end{figure}

The basic building block of an ANN is the \textit{neuron}. Information is passed as inputs to the neuron, which processes them and produces an output. The output is typically a simple mathematical function of the inputs. The output of an ANN can be interpreted as a Bayesian {\em a posteriori} probability that models the likelihood of membership class on the basis of input parameters \citep{gis90, ric91}. Hereafter we refer to such a probability as $L$. The power of ANNs comes from assembling many neurons into a network. The network is able to model very complex behavior from input to output. ANNs exist in many different models and architectures. Because of the relatively low complexity of our data, we decided to use a simple neural model known as \textit{Feed Forward MultiLayer Perceptron} and in particular a two-layer feed-forward network (2LP), which is probably the most widely used architecture for practical applications of neural networks. It consists of a layer of input neurons, a layer of ``hidden'' neurons and a layer of output neurons. In such an arrangement each neuron is referred to as a \textit{node}. The nodes in a given layer are fully connected to the nodes in the next layer by links. For each input pattern, the network produces an output pattern, compares the actual output with the desired one and computes an error. The error is then reduced by an appropriate quantity adjusting the weights associated to each link through a specific learning algorithm.  This process continues until the error is minimized. Fig.~\ref{ann} shows a schematic design of such a network.

In $\gamma$-ray astronomy, ANNs are often used for such applications as background rejection, though other techniques (e.g. classification trees) are also used for such purposes. In recent years ANNs were also used for classifying {\em Fermi}-LAT unassociated sources \citep{doe14}. This technique uses identified objects as a training sample, learning to distinguish each source class on the basis of parameters that describe its $\gamma$-ray properties. By applying the algorithm to unknown objects, such as the unclassified sources, it is possible to quantify their probability of belonging to a specific source class. There are different packages available to perform an ANN analysis (e.g. \textit{MATLAB Neural Network Toolbox}\footnote{{\tt http://www.mathworks.com/products/neural-network/}} or \textit{PyBrain}\footnote{{\tt http://pybrain.org}}), but we decided to develop our own 2LP algorithms to address our specific problem. We wrote our algorithms in \textit{Python} programming language\footnote{{\tt http://www.python.org}}. Our choice gives us a number of advantages. First of all our ANN does not work as a ``\textit{black box}'', which is a typical problem of any available ANN package for which the learning process is always unknown. Since we have implemented our algorithms, we can examine step by step how our network is learning to distinguish 3FGL source classes.

To date, \citet{ack12, lee12, mir12, has13, doe14, park16} have explored the application of machine learning algorithms to source classification, based on some $\gamma$-ray observables, showing that there is much to be gained in developing an automated system of sorting (and ranking) sources according to their probability of being a specific source class. The present work differs from these in applying the technique to different types of blazars rather than trying to separate AGN in general from other source classes.

We tuned a number of ANN parameters to improve the performance of the algorithm.
We renormalized all input parameters between 0 and 1 to minimize the influence of the different ranges. We used a hyperbolic tangent function as activation function connected to each hidden and output nodes. The outputs were renormalized between 0 and 1 to handle them as a probabilities of class membership. We randomly initialized the weights in the range between -1 and 1, not including any bias. The optimal number of hidden nodes was chosen through the {\em pruning} method \citep{ree93}. We used the standard back-propagation algorithm as learning method setting the learning rate parameter to 0.2. We did not add the momentum factor in the learning algorithm because it does not improve the performance of the network. We used the learning algorithm in the {\em on-line} version, in which weights associated to each link are updated after each example is processed by the network.

\subsection{Source sample and predictor parameters}

Since the aim of this work is to quantify the likelihood of each 3FGL BCU {\bf being} more similar to a BL Lac or a FSRQ, we chose all  660 BL Lacs and 484 FSRQs in the 3FGL catalog as a source sample.
This is a two-class approach, where the output \bll\ expresses the likelihood of a BCU source to belong to the BL Lac source class and \fsrq$=1-$\bll\ to the FSRQ one. Because our interest is only in blazars, we do not expect any contribution to the BCU sample from other extragalactic source classes, and thus we did not estimate their contamination in our analysis. We encoded the output of the associated blazars so that \bll\ is 1 if the known object is a BL Lac, and \bll\ is 0 if it is a FSRQ.

Following the standard approach, we randomly split the 3FGL blazar sample into 3 subsamples: the training, the validation and the testing one. The training sample is used to optimize the network and classify correctly the encoded sources. The validation sample is used to avoid over-fitting during the training. This is not used for optimizing the network, but during the training session it monitors the generalization error. The learning algorithm is stopped at the lowest validation error. The testing sample is independent both of the training and validation ones and was used to monitor the accuracy of the network. Once all optimizations were made, the network is applied to the testing sample, and the related error provides an unbiased estimate of the generalization error. We chose a training sample as large as possible ($\sim$ 70\% of the full sample) while keeping the other independent samples homogeneous ($\sim$ 15\% {\bf for} each one). Since we used an {\em on-line} version of the learning algorithm, we decided to shuffle the training sample after the full training  sample was used once to optimize the network. This choice allowed us to maintain a good generalization of our network.

Because we want to distinguish BL Lacs from FSRQs only on the basis of their $\gamma$-ray ECDF, we selected flux values extracted from such a distribution as predictor parameters. We included in our ANN algorithm $\gamma$-ray fluxes corresponding to 10th, 20th, 30th, 40th, 50th, 60th, 70th, 80th, 90th and 100th percentile

Our choice to use only 10 input parameters originates from a compromise between a good representation of each ECDF and a limited number of input parameters, in order to avoid problems related to upper limits associated to some bin times.

\begin{figure}\resizebox{\hsize}{!}{\includegraphics{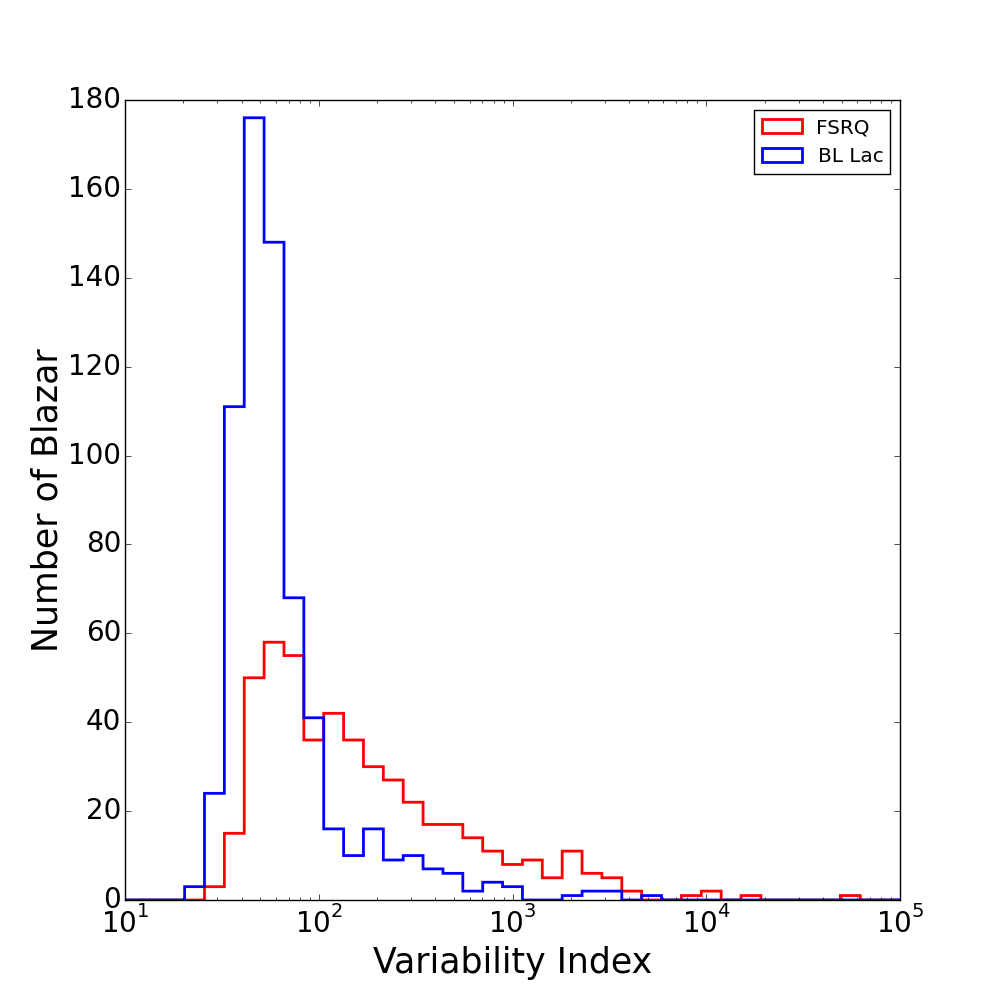}}
\caption{Variability Index distribution for 3FGL BL Lacs (blue) and FSRQs (red).}\label{var_dist}
\end{figure}

We also tested the performance of the network adding the Variability Index defined in the 3FGL catalog \citep{ace15} as an additional  parameter.
The Variability Index is a statistical parameter that tests if a $\gamma$-ray source is variable above a certain confidence level, in particular if its value is greater than $72.5$ the object is statistically variable at the 99\% confidence level. The information given by the Variability Index is more limited than the ECDF, which also provides a characterization of the variability pattern and is probably related to spectral variability during the flare state. Including the Variability Index in the algorithm did not significantly improve the performance of the network,  showing that this parameter does not add independent information in distinguishing the two blazar subclasses. Defining the importance of each input parameter as the product of the {\bf mean-square} of the input variables with the sum of the weights-squared of the connection between the variable's nodes in the input layer and the hidden layer, the Variability Index was observed to be the less important parameter. Fig.~\ref{var_dist} confirms that the distribution of Variability Index is very similar for 3FGL BL Lacs and FSRQs. Although the mean variability is higher for FSRQs, the distributions overlap strongly, making this parameter hard to use as a discriminator.

{We excluded from our analysis both $\gamma$-ray and multiwavelength spectral parameters, because the aim of this work is to develop a classification algorithm that can be efficiently applied to $\gamma$-ray sources when rigorous $\gamma$-ray spectra or multiwavelength information is missing. Since the best way yet found to single out  BL  Lacs from FSRQs is to analyse their spectral energy distribution \citep{Ackermann2015}, we used multiwavelength spectral information to validate our algorithm, comparing the distribution of BL Lac and FSRQ candidates with known ones as discussed in Section~\ref{sec:results},~\ref{sec:optical} and~\ref{sec:radio}.

As a result of these choices, our feed-forward 2LP is built up of 10 input nodes, 6 hidden nodes and 2 output nodes.

\subsection{Optimization of the algorithm and classification thresholds}

At the end of the learning session, the ability of the algorithm to distinguish BL Lacs from FSRQs is optimized, and for each blazar produces a likelihood of its membership class. Fig.~\ref{ann_bll_fsrq} shows the likelihood distribution applied to the testing sample. The distribution clearly shows two distinct and opposite peaks for BL Lac (blue) and FSRQ (red), the former at \bll $\sim 1$ while the latter at \bll\ $\sim 0$. Since the testing sample was not used to train the network, the distribution shows the excellent performance of our algorithm in classifying new BL Lacs and FSRQs.

\begin{figure}\resizebox{\hsize}{!}{\includegraphics{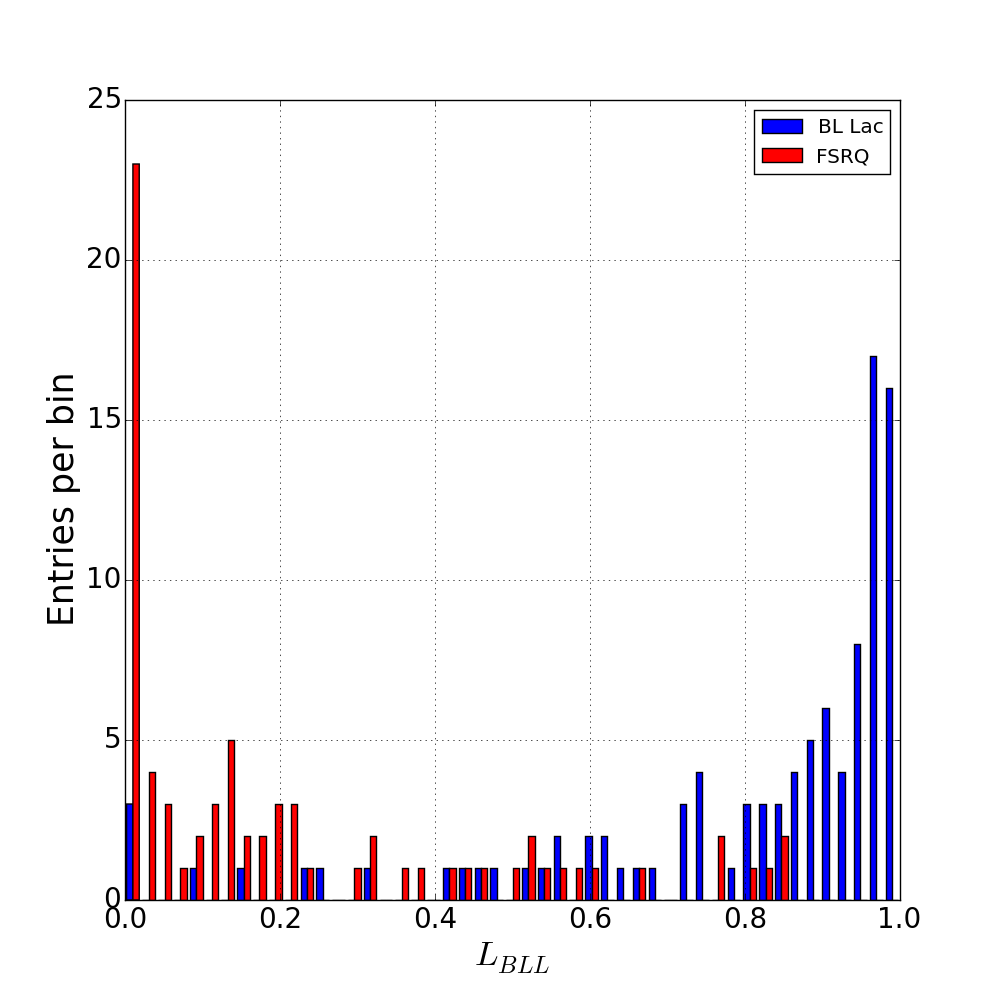}}
\caption{Distribution of the ANN likelihood to be a BL Lac candidate for 3FGL BL Lacs (blue) and FSRQs (red) in the testing sample. The distribution of the likelihood to be a FSRQ candidate (\fsrq) is $1-$\bll.}\label{ann_bll_fsrq}
\end{figure}

We defined two classification thresholds to label BCUs as BL Lac or FSRQ candidates. Our thresholds are based on the optimization of the positive association rate (\emph{precision}), which is defined as the fraction of true positives with respect to the objects classified as positive, of $\sim 90\%$. The classification threshold of \bll$>0.566$ identifies BLL candidates, while threshold \fsrq$>0.770$ identifies FSRQ candidates. Another parameter useful to characterize the performance of our classification algorithm is the \emph{sensitivity}, defined as the fraction of objects of a specific class correctly classified as such. According to this definition, the threshold for BL Lac classification is characterized by a sensitivity of $\sim 84\%$, while we get a sensitivity of ~69\% for FSRQs. The precision and sensitivity of our classification algorithm help us to predict the completeness and the fraction of spurious sources in the list of BL Lac and FSRQ candidates. Thresholds defined on the basis of high \emph{precision} are useful to select the best targets to observe with ground telescopes, optical or Cherenkov, to unveil their nature, while high \emph{sensitivity} gives us an idea of how many BL Lacs and FSRQs remain to be identified in the 3FGL BCU sample.
In the end, according to our classification thresholds, the expected false negative rate (\emph{misclassification}) is $\sim5\%$ for BL Lacs and $\sim12\%$ for FSRQs.
\emph{Sensitivity}, \emph{misclassification} and \emph{precision}
reveal that the FSRQ $\gamma$-ray ECDF is broader and more contaminated than the BL Lac one, as we expected from Fig.~\ref{<Fig.1>}. 
The combination of high precision rate and low misclassification rate indicates a very high performance of our optimized network.

\subsection{Selecting the most promising HSP candidates}

Although the ECDF of HSPs are not clearly separated from those of ISPs and LSPs, we developed a new ANN algorithm to select the best HSP candidates among BCUs, in order to 

optimize observations by VHE facilities. Following the procedure described in the previous sections we chose as a source sample all 289 HSPs and the 824 non-HSPs identified by their spectral energy distribution. We used as predictor variables the same ECDF parameters used to classify BLLs and FSRQs. The new feed-forward 2LP is built up of 10 input nodes, 5 hidden nodes and 2 output nodes.

Fig.~\ref{ann_hsp} shows the optimized networks applied to a testing sample that represents $15\%$ of the full sample. The distribution reveals a peak at low $L_{HSP}$ for non-HSP and a nearly flat distribution for HSP sources, showing the optimized network was not able to clearly classify HSPs on the basis of ECDF as expected. Defining a classification threshold of $L_{HSP}>0.891$ so that the \emph{precision} rate is $\sim90\%$, we are able to discover the best HSP candidates. According to this definition, the \emph{sensitivity} of our algorithm is just $4.5\%$ while the fraction of non-HSPs erroneously classified as HSP candidates is very low ($<1\%$). This result shows that only a very small fraction of HSPs can be separated from non-HSPs by this method. We name all the BCUs in this region as \emph{Very High Confidence} (VHC) HSP candidates. All the blazars in this area are BL Lacs. The only FSRQ characterized by a higher $L_{HSP}$ value, $\sim0.85$, is 3FGL J1145.8+4425. This means that all the VHC HSP candidates will also be VHC BL Lac candidates.
In addition, we decided to define a less conservative classification threshold ($L_{HSP}>0.8$) in order to increase the number of targets to observe with VHE telescopes at the expense of a smaller precision ($\sim75\%$). In this way the \emph{sensitivity} increases to $\sim15\%$ and the misclassified non-HSP remains very low ($\sim2\%$). We label BCU characterized by a $L_{HSP}$ greater than such a classification threshold as High Confidence (HC) HSP candidates.}

\begin{figure}\resizebox{\hsize}{!}{\includegraphics{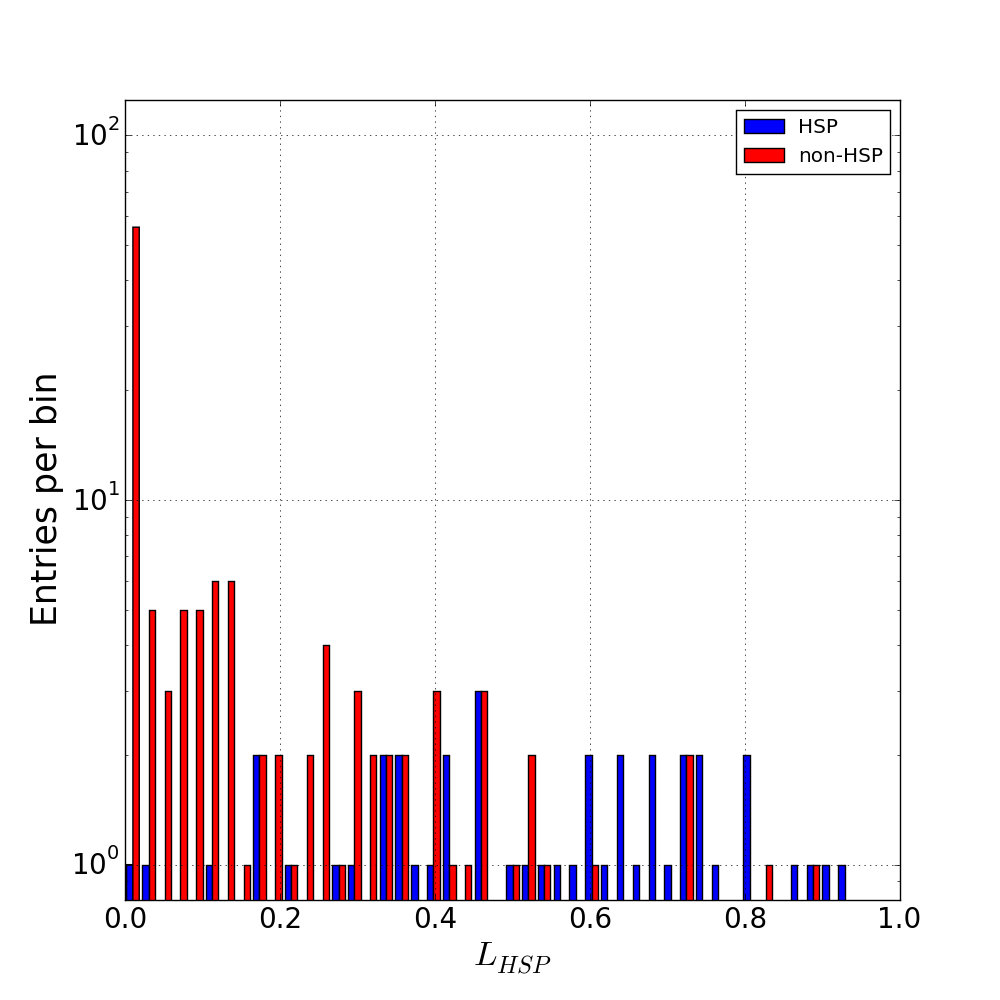}}
\caption{Distribution of the ANN likelihood to be a HSP candidate for HSP (blue) and non-HSP (red) in the testing sample.}\label{ann_hsp}
\end{figure}

\section{ANN Results and validation}\label{sec:results} 

In this section we first discuss the results of our optimized ANN algorithm at classifying BL Lac and FSRQ candidates among 3FGL BCU sources. Then we validate our statistical method comparing the PowerLaw Index distribution of known BL Lacs and FSRQs with that of our best candidates. Then we  analyze the performance of our algorithm based on ECDF with respect to the other $\gamma$-ray parameters usually used to classify blazars, such as PowerLaw Index and Variability Index. In the end we  discuss the results on the identification of the most promising HSP candidates.

\begin{figure}\resizebox{\hsize}{!}{\includegraphics{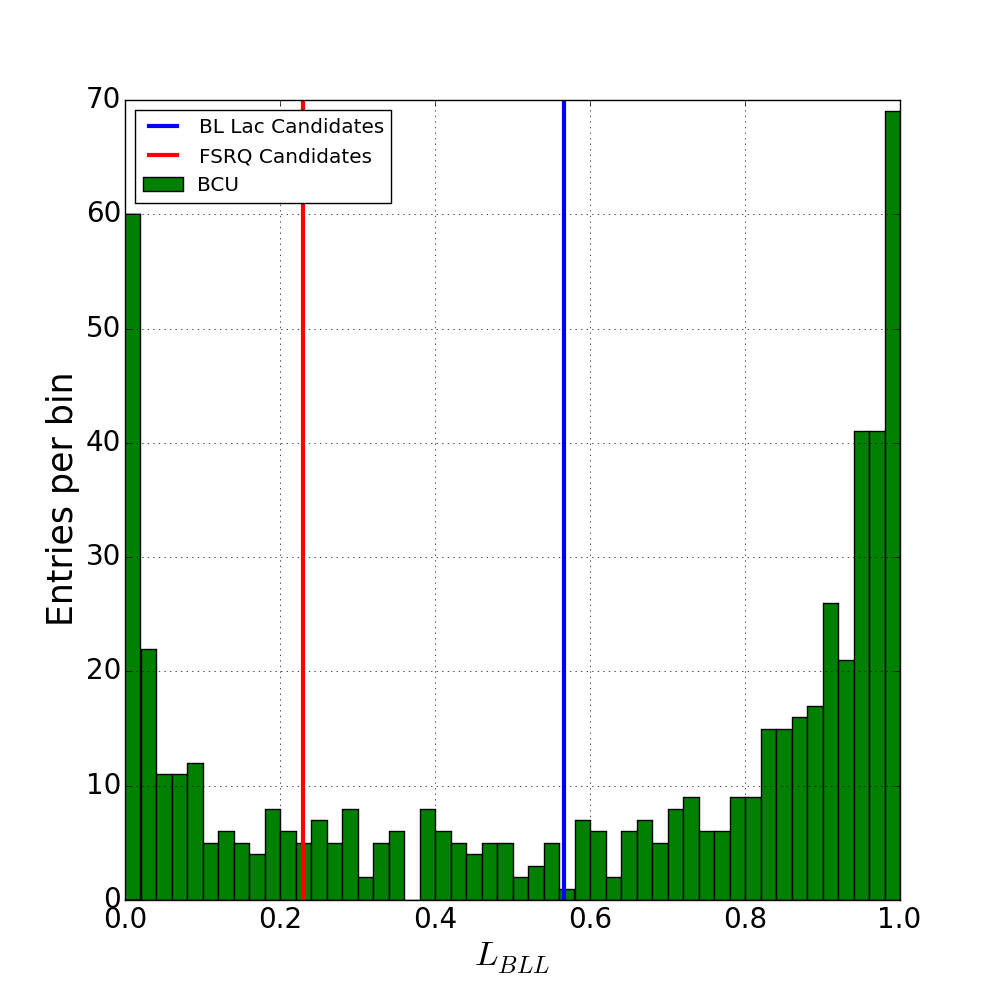}}
\caption{Distribution of the ANN likelihood of 573 3FGL BCU to be BL Lac candidates. Vertical blue and red lines indicate the classification thresholds of our ANN algorithm to label a source as BL Lac or FSRQ respectively as described in the text.}\label{ann_agu}
\end{figure}

Applying our optimized algorithm to 573 3FGL BCUs we find that 342 are classified as BL Lac candidates ($L_{BLL} > 0.566$), 154 as FSRQ candidates ($L_{FSRQ} > 0.770$) and 77 remain unclassified. Hereafter we will define as  \bfgl\ and \ffgl\  blazars classified in the 3FGL catalog, while as \bann\ and \fann\  BCUs classified by ANN and \bcann\  BCUs that remain uncertain. The likelihood distribution of BCUs membership class is shown in Fig.~\ref{ann_agu} and such a distribution reflects very well those of \bfgl\ and \ffgl\ in the testing sample (see Fig.~\ref{ann_bll_fsrq}) as we expect for a well-built classification algorithm. Taking into account {\em precision} and {\em sensitivity} rates, our optimized algorithm predicts that there are about 365 BL Lacs and about 200 FSRQs to be still identified. This prediction is rather interesting, because at present the fraction of \bfgl\  is $\sim1.4$ times that of \ffgl\ while a larger fraction ($\sim1.8$) of BL Lacs to be identified is expected by our analysis.

\begin{figure}\resizebox{\hsize}{!}{\includegraphics{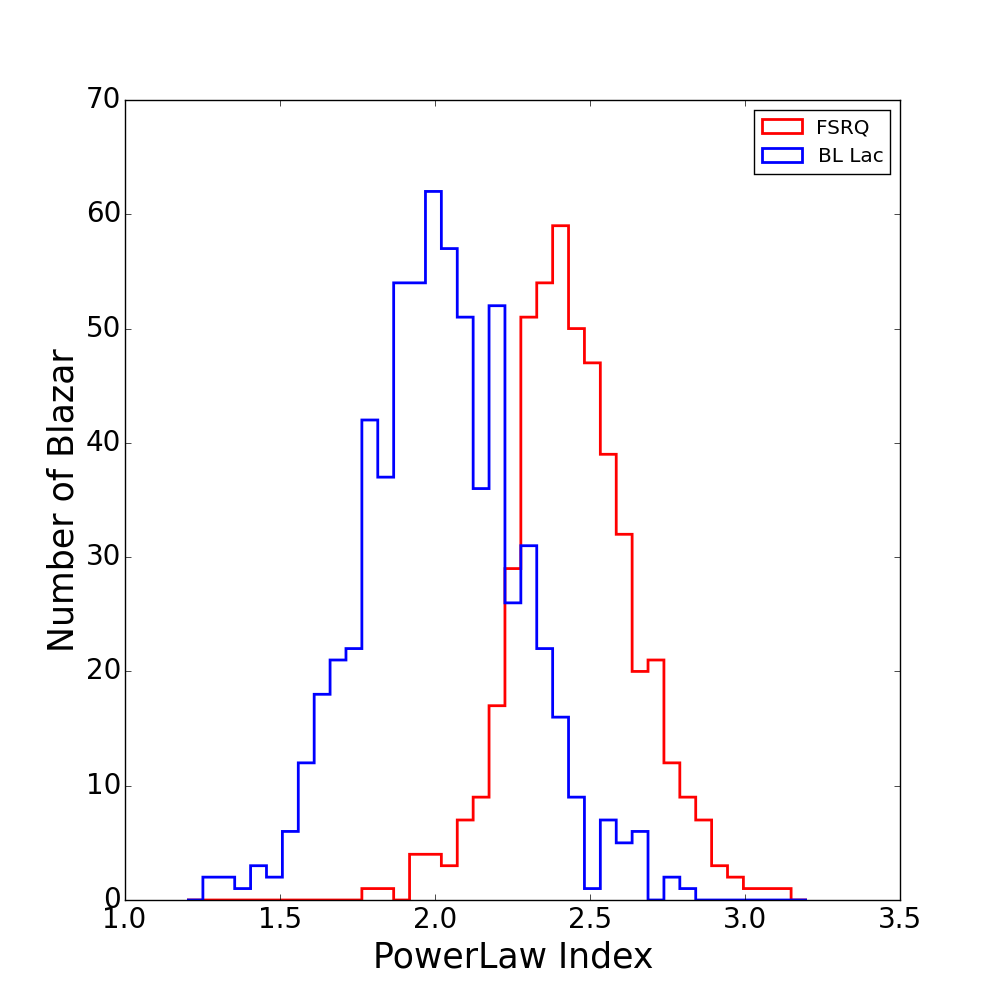}}
\caption{PowerLaw Index distribution for 3FGL BL Lacs (blue) and FSRQs (red).}
\label{spec_distribution}
\end{figure}

\begin{figure*}
\label{pi_dist}
\begin{center}
\includegraphics[width=0.48\textwidth]{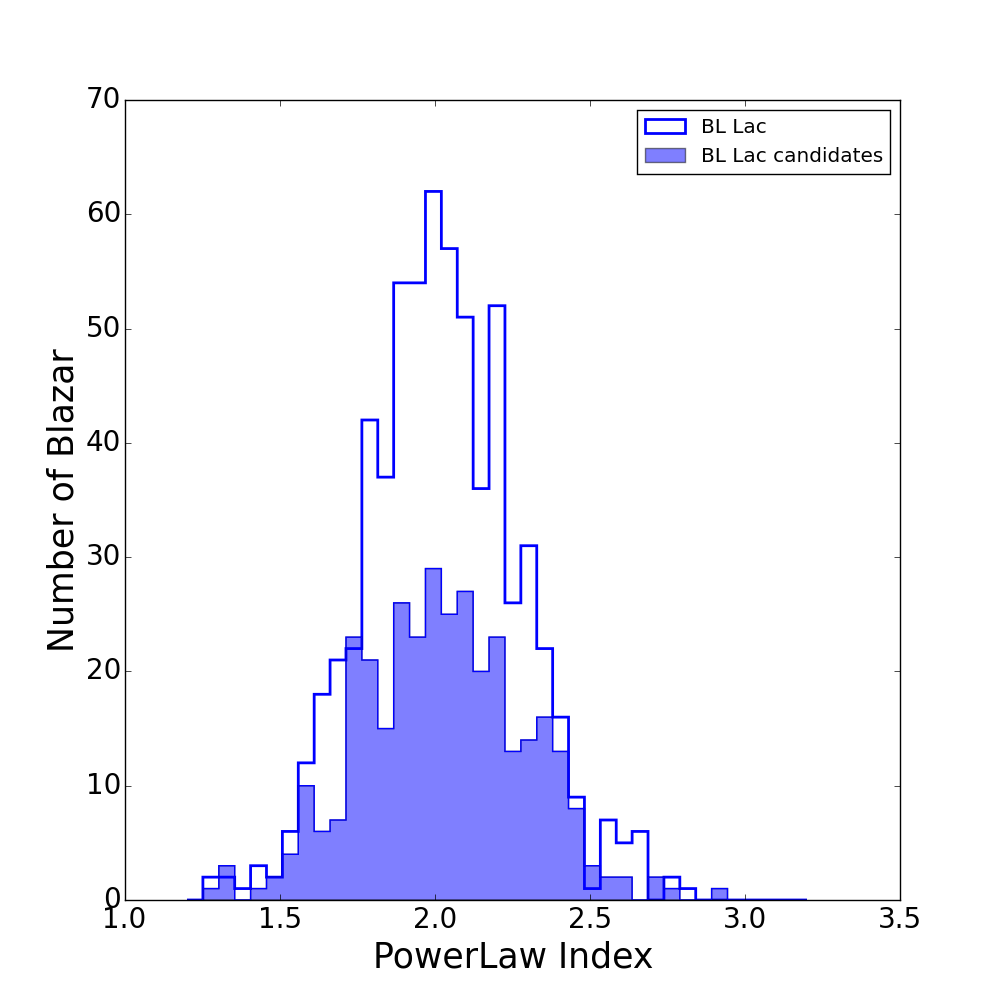}
\includegraphics[width=0.48\textwidth]{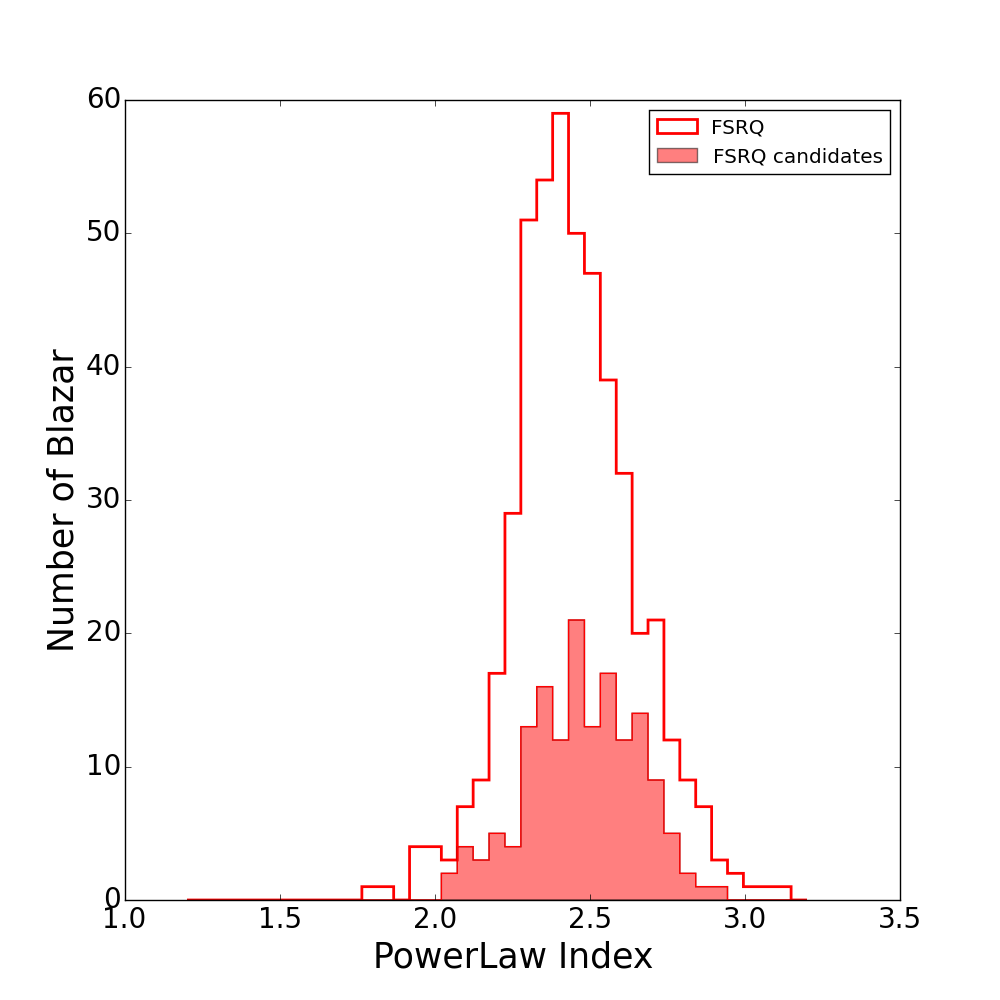}
\caption{PowerLaw Index distribution for the blazars classified through the ANN method (filled histograms) in comparison to the previously classified blazars. Left: BL Lacs; right: FSRQs.}
\end{center}
\end{figure*}

After the launch of the {\em Fermi} observatory it was discovered that BL Lacs and FSRQs are characterized by different $\gamma$-ray spectral properties. The former usually show harder spectra than the latter \citep{Ackermann2015}. Fitting 3FGL blazars assuming a power-law spectral model we observe that the best-fit photon spectral index (in 3FGL named \textit{PowerLaw Index}) distribution is rather dissimilar for the two subclasses as shown in Fig.~\ref{spec_distribution}. The PowerLaw Index distribution {\bf mean values} and standard deviations are $2.02\pm0.25$ and $2.45\pm0.20$ for BL Lacs and FSRQs respectively, making this observable one of the most powerful $\gamma$-ray parameters to distinguish the two blazar subclasses. Since we did not include this parameter in our algorithm, we compared the PowerLaw Index distribution for \bann\ and \fann\ with what we know from already classified objects to test the performance of our algorithm and to validate it. Fig.~\ref{pi_dist} shows in the left panel the PowerLaw Index distributions for BL Lacs while in the right one for the FSRQs.  Such distributions are in good agreement, confirming the accuracy of our classification algorithm. The PowerLaw Index distribution means and standard deviations are $2.02\pm0.27$ and $2.48\pm0.18$ for \bann\ and \fann\ respectively as expected. Moreover almost all sources classified through the BFLaP-ANN method are within the PowerLaw Index distribution range associated to their blazar subclass.

\begin{figure}\resizebox{\hsize}{!}{\includegraphics{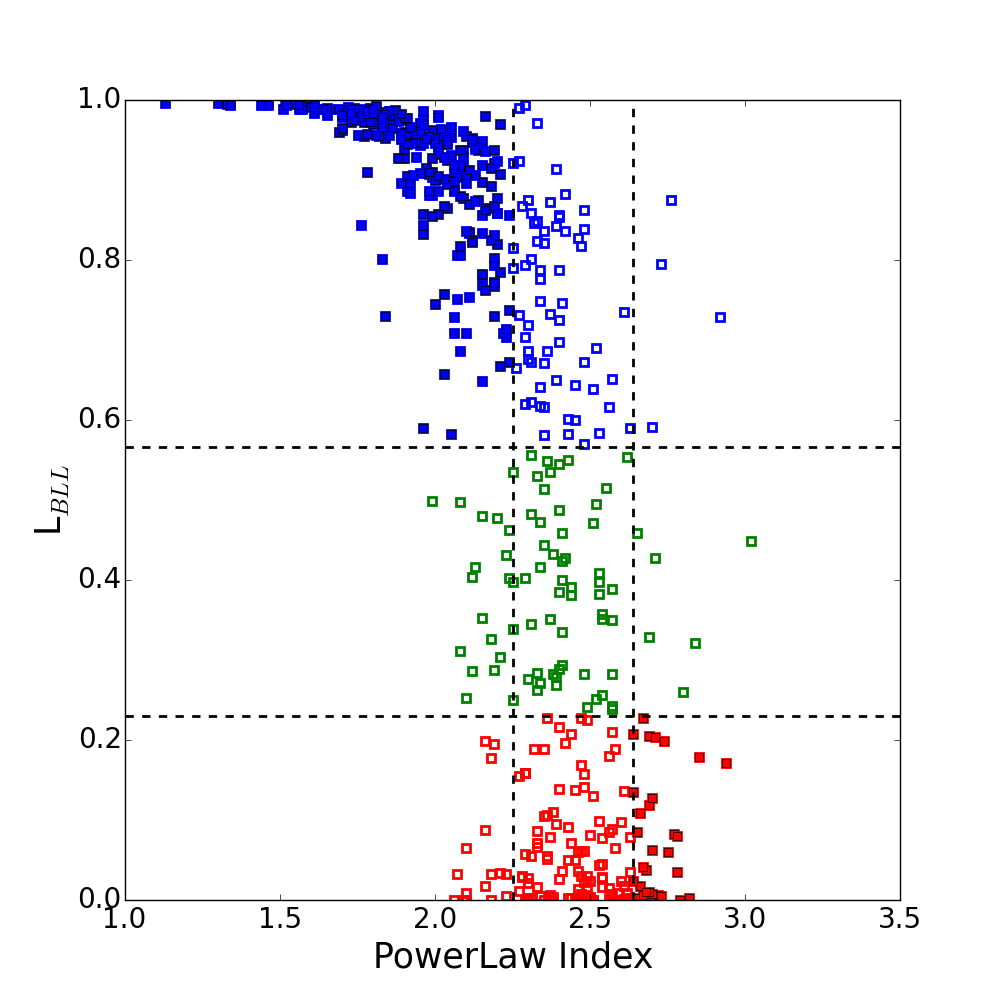}}
\caption{ANN likelihood against PowerLaw Index distributions. Colors indicate the classification proposed by the ANN method: blue for BL Lacs, red for FSRQs, and green for still uncertain objects. Filled symbols indicate the sources for which the PowerLaw Index indicates a matching classification.}
\label{ann_vs_pi}
\end{figure}

An effective way to evaluate the power of our method is to compare ANN predictions for distinguishing blazar subclasses based on B-FlaP information with those found by a simple analysis of $\gamma$-ray spectral or timing properties.
Analyzing the PowerLaw Index distribution shown in Fig.~\ref{spec_distribution} we can define two classification thresholds to separate BL Lacs from FSRQs with a degree of purity equal to what we used for ANN thresholds, 90\%. According to this hypothesis all blazars characterized by a PowerLaw Index $<2.25$ or $>2.64$ will be classified as BL Lac and FSRQ candidates respectively with a precision rate of 90\%. All blazars with an intermediate value will remain unclassified owing to high contamination. Fig.~\ref{ann_vs_pi} shows the PowerLaw Index distribution against the ANN likelihood to be a BL Lac of all 3FGL BCUs. Vertical and horizontal dashed lines indicate classification thresholds defined for the two distributions to single out BL Lacs from FSRQs. Comparing the two predictions we observe they agree for $\sim63\%$ of BCUs (blocks along the diagonal from top left to bottom right), while disagree only for $\sim3.5\%$ (top right and bottom left blocks). As a key result we observe that ANN method based on B-FlaP is able to provide a classification for $\sim30\%$ of BCUs remaining uncertain on the basis of their spectra (top and bottom central blocks) while the opposite occurs only for $\sim3.5\%$ of BCUs. This comparison highlights the power of our analysis with respect to the standard one based on spectral information.

\begin{figure}\resizebox{\hsize}{!}{\includegraphics{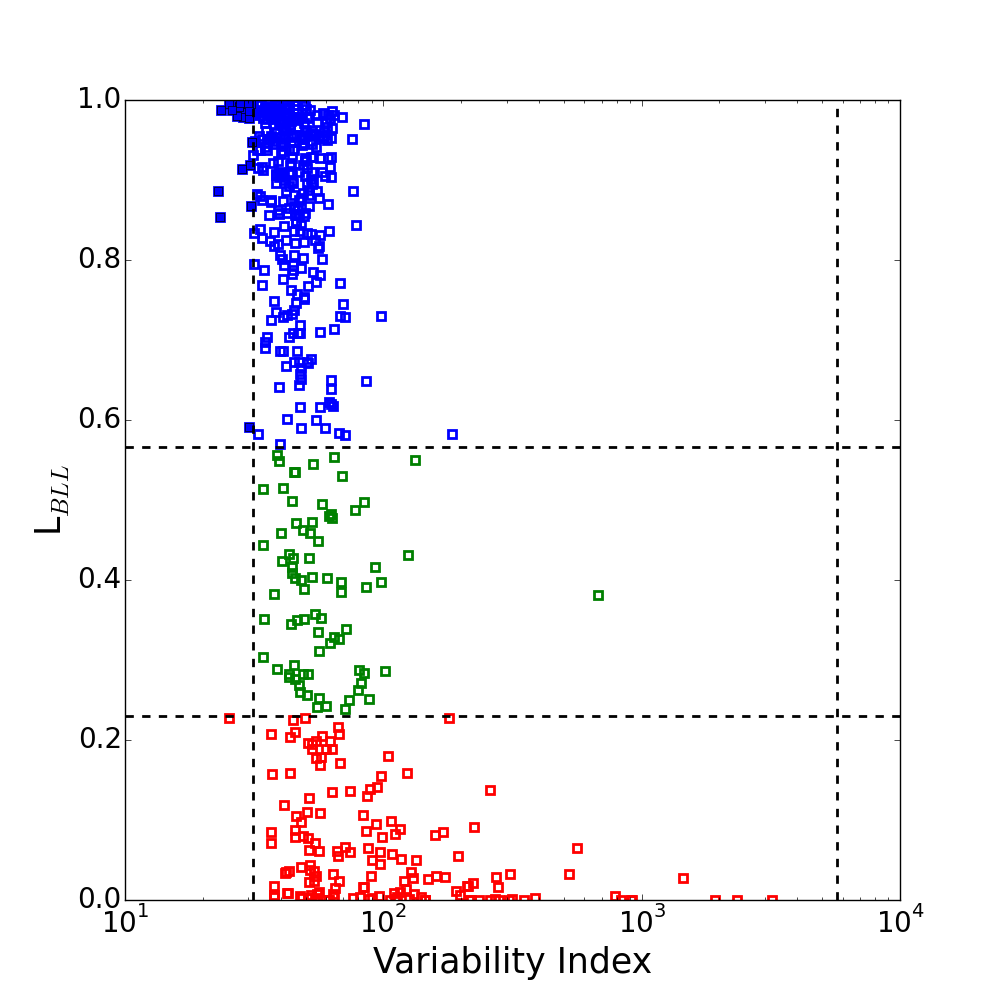}}
\caption{ANN likelihood against Variability Index distributions as described in Fig.~\ref{ann_vs_pi}}
\label{ann_vs_vi}
\end{figure}

To be thorough we followed the same approach to compare ANN predictions based on B-FlaP with those obtained by Variability Index. As discussed in the previous Section, we expect this parameter is not efficient at distinguishing blazar subclasses so that we did not include it in our analysis. We defined two classification thresholds as before from the Variability Index distribution (see Fig.~\ref{var_dist}) so that blazars with a value smaller than 31 are classified as BL Lac candidates while those with a value larger than 5710 are FSRQ candidates in agreement with the 90\%  precision criterion. These areas are very small because the overlap in the Variability Index distribution is very large. As shown in Fig.~\ref{ann_vs_vi}, the two methods agree only for $\sim17\%$ of BCUs and disagree for $\sim0.2\%$. No BCU classified by the Variability Index remains uncertain with ANN, while for a very large fraction, $\sim83\%$, ANN is able to provide a classification where the Variability Index is not. This analysis clearly shows Variability Index is {\bf not effective} at classifying blazar subclasses as we expect, and it must be replaced by the more robust B-FLaP for this purpose.

\begin{figure}\resizebox{\hsize}{!}{\includegraphics{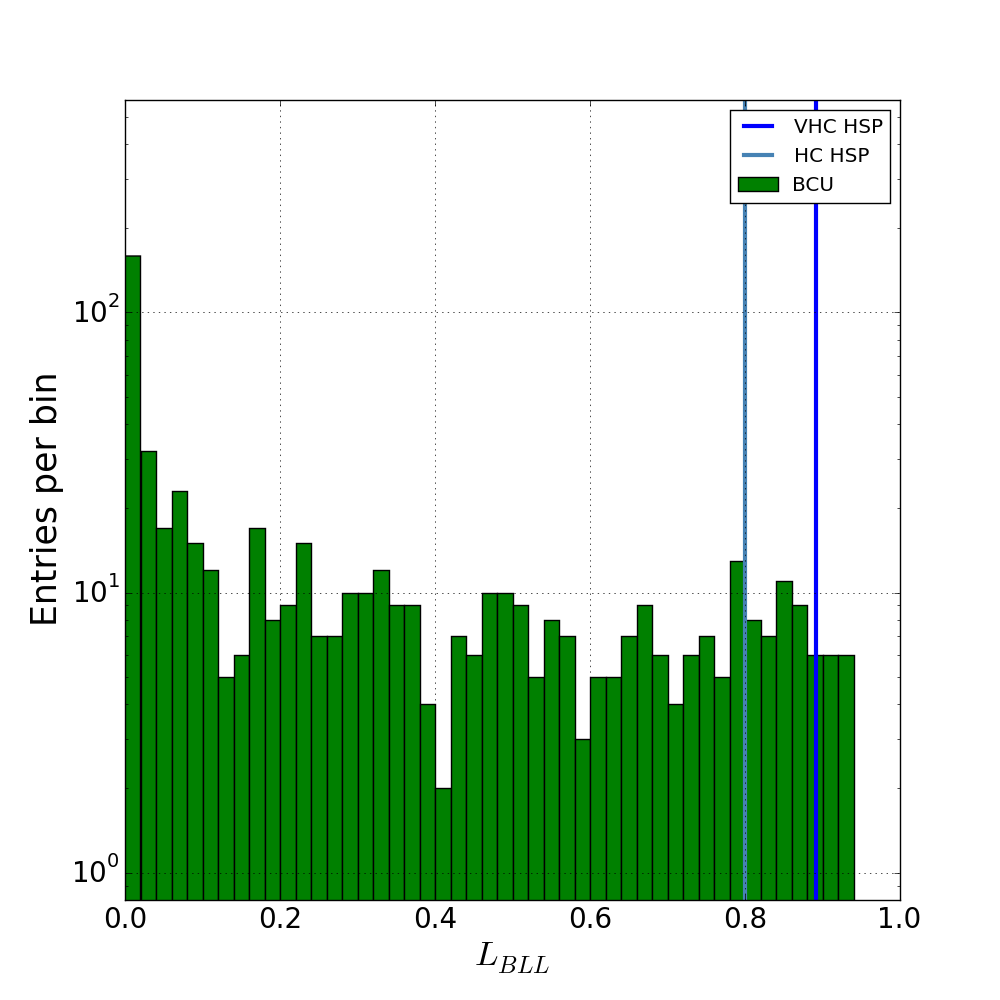}}
\caption{Distribution of the ANN likelihood of 573 3FGL BCUs to be HSP candidates. Vertical blue and steel blue lines indicate the classification thresholds of our ANN algorithm to identify a source as Very High Confidence or High Confidence HSP respectively as described in the text.}\label{ann_agu_hsp}
\end{figure}

In the end, applying our algorithm optimized to select the most promising HSPs among  573 3FGL BCUs, we can single out 15 VHC HSP candidates ($L_{HSP}>0.891$) and 38 HC ones ($L_{HSP}>0.8$) for a total of 53 very interesting targets to be observed through Very High Energy telescopes. Fig.~\ref{ann_agu_hsp} plots the likelihood distribution of BCUs. Such a distribution reflects very well those of the entire testing sample (see Fig.~\ref{ann_hsp}) showing a nearly flat distribution at high $L_{HSP}$ values related to a large overlap between HSPs and non-HSPs in the B-FlaP parameter space. We compared our predictions with those found by the 3LAC catalog on the basis of the study of broadband Spectral Energy Distributions (SED) collected from all data available in the literature. The SED classification  \citet{Ackermann2015} 
is based on the estimation of the synchrotron peak frequency $\nu^S_{peak}$ value extracted from a 3rd-degree polynomial fit
of the low-energy hump of the SED. Out of 15 VHC HSPs, 11 ($\sim73\%$) are classified as HSPs on the basis of their broadband SED and 4 ($\sim28\%$) remain unclassified. Out of 38 HC HSPs, 22 ($\sim58\%$) are classified as HSPs, 8 ($\sim21\%$) are classified as non-HSPs and 8 ($\sim21\%$) remain unclassified by their broadband SED. To conclude, classifications agree for $\sim63\%$ of most promising HSPs selected by ANN, validating the efficiency of our algorithm; they disagree for $\sim15\%$, in agreement with the expected contamination rate; and for the remaining $\sim22\%$ ANN provides a classification as most promising HSPs, while the SED is not rigorous enough or available.

\begin{table}
\caption{ \bfgl , \ffgl , \bcfgl,   and the new classification of blazars after B-FlaP ANN  analysis. }
\label{bcuanntb}
\begin{center}
\begin{footnotesize}
\begin{tabular}{lcr}
\hline
\hline
\\[-1ex]
{\bf \bfgl} & {\bf \ffgl} & {\bf \bcfgl} \\ [1ex]
660(38.4$\%$) & 484(28.2$\%$) & 573(33.4$\%$)\\ [2ex]
\hline
\\[-1ex]
{\bf \btann} & {\bf \ftann} & {\bf \bctann} \\ [1ex]
1002(58.3$\%$) &638(37.2$\%$)& 77(4.5$\%$) \\[2ex]
\hline
\end{tabular}
\end{footnotesize}
\end{center}
\end{table}

\section {B- FlaP Classification List}\label{sec:class} 

Two of the main goals of our examination are to classify 3FGL BCUs as BL Lac or FSRQ candidates and to identify the most promising BCUs to target in VHE observations.

We used an innovative method to extract useful information. We investigated for the first time the distribution of blazars in the ECDF of $\gamma$-ray flux parameter space, and we applied an advanced machine learning algorithm as ANN to learn to distinguish BL Lacs from FSRQs and to recognize the most likely HSP candidates. The power of our approach was tested in the previous Section, and we present a summary of our results in Table~\ref{bcuanntb}.

The full table of individual results, available online, contains the classification of BCUs listed in the 3FGL {\it Fermi}-LAT as the key parameter. 
 We provide for each 3FGL BCU the ANN likelihood (\emph{L}) to be a BL Lac or a FSRQ, and the predicted classification  according to the defined classification thresholds.  We label the most promising HSP candidates, splitting these objects into High Confidence HSPs and Very High Confidence HSPs in agreement with their likelihood to be an HSP-like source.
Table~\ref{class_list} shows a portion of these results, the full table being available electronically from the journal.

{\bf \begin{table*}
\caption{Classification List of 3FGL BCUs -- sample. The table is published in its entirety in the electronic edition of the article. 
The columns are: 3FGL Name, Galactic Latitude and Longitude ($b$ and $l$), the ANN likelihood to be classified as a BL Lac ($L_{BLL}$) and a FSRQ ($L_{FSRQ}$), the predicted classification and the most promising HSP candidates labeled as Very High C. or High C., where C. is for {\em Confidence}}
\label{class_list}
\begin{center}
\begin{footnotesize}
\begin{tabular}{lcccccc}
\hline
\hline
3FGL Name 	&  b ($^{\circ}$) & l ($^{\circ}$) & \bll  &  \fsrq  &  Classification  &	 HSP Candidates\\
\hline

J0002.2--4152	&	--72.040	&	334.320	&	0.877	&	0.123	&	BL Lac	&		\\
J0003.2--5246	&	--62.820	&	318.940	&	0.976	&	0.024	&	BL Lac	&      		\\
J0003.8--1151	&	--71.080	&	84.660	&	0.952	&	0.048	&	BL Lac	&		\\
J0009.6--3211	&	--79.570	&	0.880	&	0.859	&	0.141	&	BL Lac	&		\\
J0012.4+7040	&	8.140	&	119.620	&	0.022	&	0.978	&	FSRQ	&		\\
J0014.6+6119	&	--1.270	&	118.540	&	0.896	&	0.104	&	BL Lac	&		\\
J0015.7+5552	&	--6.660	&	117.890	&	0.835	&	0.165	&	BL Lac	&		\\
J0017.2--0643	&	--68.150	&	99.510	&	0.953	&	0.047	&	BL Lac	&		\\
J0019.1--5645	&	--59.890	&	311.690	&	0.650	&	0.350	&	BL Lac	&		\\
J0021.6-6835	&	--48.580	&	306.730	&	0.459	&	0.541	&	Uncertain	&		\\
J0028.6+7507	&	12.300	&	121.400	&	0.749	&	0.251	&	BL Lac	&	\\
J0028.8+1951	&	--42.540	&	115.610	&	0.602	&	0.398	&	BL Lac	&	\\
J0030.2--1646	&	--78.570	&	96.580	&	0.981	&	0.019	&	BL Lac	&	High C.\\
J0030.7--0209	&	--64.580	&	110.690	&	2.35e--04	&	1.000	&	FSRQ	&	\\
J0031.3+0724	&	--55.120	&	114.190	&	0.984	&	0.016	&	BL Lac	&	\\
\hline
\end{tabular}
\end{footnotesize}
\end{center}
\end{table*}}

\section{Multiwavelength investigation}\label{sec:multiwave}

\subsection{Optical data}
\label{sec:optical}

Ultimately, the classification of a blazar depends on spectroscopy, especially optical spectroscopy to identify redshift and the presence or absence of lines.  In order to assess the reliability of the \bflapann\ method in the identification of the various blazar classes, we carried out optical spectroscopic analysis of a sample of targets listed as BCUs in 3FGL, for which we had a classification likelihood. Spectral data were obtained both by combining the public products of the 12$^{\rm th}$ data release of the Sloan Digital Sky Survey \citep[SDSS DR12,][]{Alam15} and of the 2$^{\rm nd}$ data release of the 6dF Galaxy Redshift Survey \citep[6dFGRS DR2,][]{Jones04, Jones09}, as well as by direct observations performed with the 1.22m and the 1.82m telescopes of the Asiago Astrophysical Observatory.\footnote{Observatory website at \url{http://www.dfa.unipd.it/index.php?id=305}}.

The selection of targets for spectroscopic analysis is affected by the possibility to associate the low energy counterpart within the positional uncertainty of the $\gamma$-ray source. 
Because of the verified correlation of radio flux and $\gamma$-ray flux \citep{Ghirlanda,Ackermann2011} we chose the targets for spectroscopic observations by looking for coincident emission at these frequencies. 
The typical positional uncertainties of a few arc seconds achieved by radio and X-ray instruments can constrain the source position on the sky better than the $\gamma$-ray detection and, therefore, greatly reduce the number of potential counterparts. When the candidate counterpart turned out to be covered by a spectroscopic survey, we analyzed the corresponding spectrum. If, on the contrary, it was not covered by a public survey, but it was still bright enough to be observed with the Asiago instruments (typically operating below the visual magnitude limit of $V \leq 18$ in spectroscopy), we carried out specific observations.

The observational procedure involved exposures of each target and standard star, immediately followed by comparison lamps. The spectroscopic data reduction followed detector bias and flat field correction, wavelength calibration, flux calibration, cosmic rays and sky emission subtraction. All the tasks were performed through standard IRAF tools\footnote{\url{http://iraf.noao.edu/}}, customized into a proper reduction pipeline for the analysis of long slit spectra obtained with the specific instrumental configuration of the telescopes. At least one standard star spectrum per night was used for flux calibration, while the extraction of mono-dimensional spectra was performed by tracking the centroid of the target along the dispersion direction and choosing the aperture on the basis of the seeing conditions. The sky background was estimated in windows lying close to the target, in order to minimize the effects of non-uniform sky emission along the spatial direction, while cosmic rays were identified and masked out through the combination of multiple exposures of the same target. The targets for which we obtained spectral data are listed in Table~\ref{tabSpectra}. The sample is described with reference to the 3LAC terminology which divides BCUs into three sub-types: 
\begin{itemize}
\item{BCU I  has a published optical spectrum but not sensitive enough for a classification as an FSRQ or a BL Lac}
\item{ BCU II is lacking an optical spectrum but a reliable evaluation of the SED synchrotron-peak position is possible}
\item{BCU III  is lacking both an optical spectrum and an estimated synchrotron-peak position but shows blazar-like broadband emission and a flat radio spectrum.}
\end{itemize}

\begin{table*}
\caption{The sample of objects selected from the 3FGL Source Catalogue for optical observation. The table columns report, respectively, the 3FGL source name, the associated counterpart, the coordinates (right ascension and declination) of the $\gamma$-ray signal centroid, the 3LAC classification of the counterpart and the source of optical spectroscopic data. }
\label{tabSpectra}
\begin{center}
\begin{footnotesize}
\begin{tabular}{llcccr}
\hline
\hline
3FGL name & Counterpart & R.A. (J2000) & Dec. (J2000) & 3LAC class. & Data source \\
\hline
J$0040.3+4049$ & B3 0037+405           & $00:40:19.9$ & $+40:49:05$ & BCU I   & Asiago T182 \\
J$0043.7-1117$ & 1RXS J004349.3-111612 & $00:43:44.4$ & $-11:17:17$ & BCU II  & Asiago T122 \\
J$0103.7+1323$ & NVSS J010345+132346   & $01:03:45.8$ & $+13:23:31$ & BCU III & Asiago T122 \\
J$0134.5+2638$ & 1RXS J013427.2+263846 & $01:34:31.2$ & $+26:38:17$ & BCU I   & Asiago T122 \\
J$0156.3+3913$ & MG4 J015630+3913      & $01:56:22.3$ & $+39:13:52$ & BCU II  & Asiago T122 \\
J$0204.2+2420$ & B2 0201+24            & $02:04:14.9$ & $+24:20:38$ & BCU II  & Asiago T122 \\
J$0256.3+0335$ & PKS B0253+033         & $02:56:19.9$ & $+03:35:46$ & BCU II  & Asiago T122 \\
J$0339.2-1738$ & PKS 0336-177          & $03:39:12.5$ & $-17:38:42$ & BCU I   & 6dFGRS \\
J$0602.2+5314$ & GB6 J0601+5315        & $06:02:14.9$ & $+53:14:06$ & BCU I   & Asiago T122 \\
J$0708.9+2239$ & GB6 J0708+2241        & $07:08:56.9$ & $+22:39:58$ & BCU II  & Asiago T122 \\
J$0730.5-6606$ & PMN J0730-6602        & $07:30:35.0$ & $-66:06:22$ & BCU II  & 6dFGRS-DR2 \\
J$0904.3+4240$ & S4 0900+42            & $09:04:21.1$ & $+42:40:55$ & BCU II  & SDSS-DR12 \\
J$1031.0+7440$ & S5 1027+74            & $10:31:02.9$ & $+74:40:55$ & BCU I   & Asiago T182 \\
J$1256.3-1146$ & PMN J1256-1146        & $12:56:20.9$ & $-11:46:52$ & BCU I   & 6dFGRS-DR2 \\
J$1315.4+1130$ & 1RXS J131531.9+113327 & $13:15:28.6$ & $+11:30:54$ & BCU II  & Asiago T182 \\
J$1412.0+5249$ & SBS 1410+530          & $14:12:04.8$ & $+52:49:01$ & BCU I   & SDSS-DR12 \\
J$1418.9+7731$ & 1RXS J141901.8+773229 & $14:18:59.3$ & $+77:31:01$ & BCU II  & Asiago T122 \\
J$1647.4+4950$ & SBS 1646+499          & $16:47:29.5$ & $+49:50:13$ & BCU I   & Asiago T122 \\
J$1736.0+2033$ & NVSS J173605+203301   & $17:36:04.8$ & $+20:33:43$ & BCU II  & Asiago T122 \\
J$2014.5+0648$ & NVSS J201431+064849   & $20:14:33.8$ & $+06:48:36$ & BCU II  & Asiago T122 \\
\hline
\end{tabular}
\end{footnotesize}
\end{center}
\end{table*}

\begin{figure*}
\begin{center}
\includegraphics[width=0.48\textwidth]{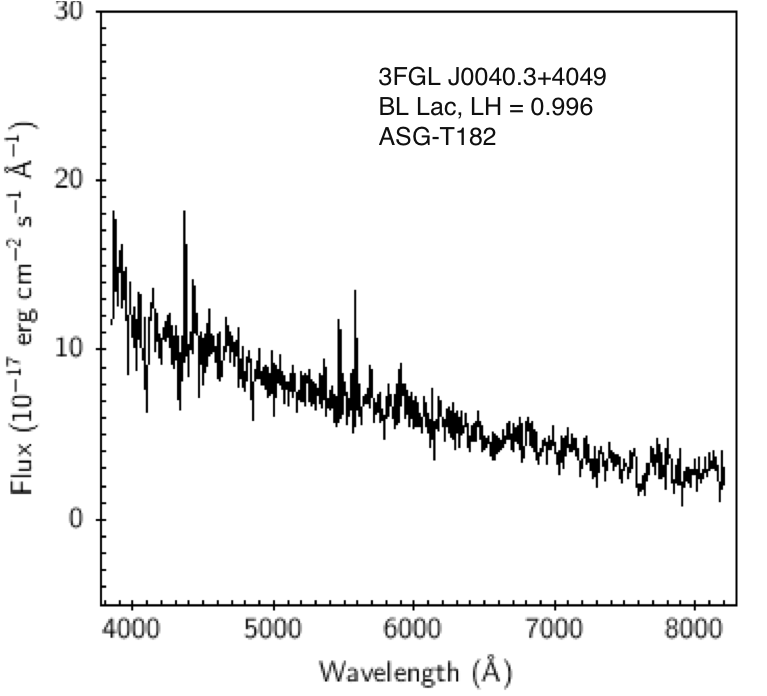}
\includegraphics[width=0.48\textwidth]{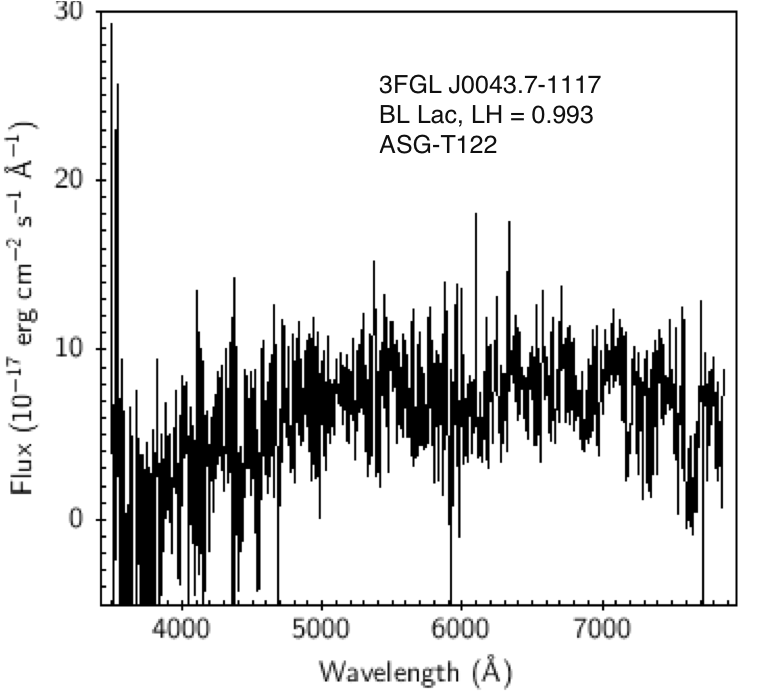}
\end{center}
\caption{Optical spectra of the low energy counterparts associated with the 3FGL blazars candidates of uncertain classification collected in this study. With the exception of 3FGL~J0156.3+3913 (\bll$= 0.024$), 3FGL~J1031.0+7440 (\bll$= 0.783$) and 3FGL~J1647.4+4950 (\bll$=0.550$) whose prominent emission lines indicate a FSRQ classification, the other sources do not show clear line emission activity and are therefore consistent with BL Lac classification, though at different levels of activity. The full set of spectra is presented in appendix.}
\label{FigOptSpec}
\end{figure*}

\subsubsection{Results}
The spectra obtained with our optical campaign are illustrated in Fig.~\ref{FigOptSpec}. Although the various targets were probably observed at different levels of activity, most of the objects located at high redshift (with $z \geq 0.1$) turned out to belong to the typical BL Lac and FSRQ blazar families. In four cases we detect clear indications of emission lines, which are not expected in BL Lac objects. These are 3FGL~J0156.3+3913, a prototypical FSRQ with $z = 0.456$ and \bll$= 0.024$, 3FGL~J0904.3+4240, a high redshift FSRQ with $z = 1.342$ and \bll$=0.673$, 3FGL~J1031.0+7440, a Seyfert 1 / FSRQ at $z = 0.122$ and \bll$= 0.783$, and 3FGL~J1647.4+4950, a Seyfert 1.9 with $z = 0.0475$ and \bll$=0.550$.

With the adopted thresholds of \bll$\geq 0.566$ to predict a BL Lac classification and \bll$\leq 0.230$ to give a FSRQ classification, these data are fully consistent with the expected 90\% precision of the method, because only 3FGL~J0904.3+4240 and 3FGL~J1031.0+7440 turn out to be misclassified (exactly 2 sources out of 20). We note, however, that the choice of more severe likelihood thresholds could easily give even more accurate results, at the obvious cost of classifying a smaller fraction of the BCU population.

\subsection{Radio data}\label{sec:radio}

Besides the different $\gamma$-ray properties and optical spectra, BL Lacs and FSRQs are also dissimilar in their radio properties. BL Lacs are generally less luminous than FSRQs, so a classification based on radio luminosity could be a useful diagnostic for BCUs. However, radio luminosity is a quantity that can only be calculated if a redshift is known -- and very often, nearly by definition, BCUs do not have an available optical spectrum suitable for the determination of $z$ (this is actually the case for $\sim 91\%$ of our BCUs). In any case, as we are going to show \citep[see also][]{2LAC,Ackermann2015}, the separation between BL Lacs and FSRQs remains rather clear also according to the flux density parameter. For this reason, we study here the radio flux density distribution of the 3FGL BL Lacs, FSRQs, and BCUs, in order to show that  (1) the classification proposed by our \bflapann\ method is in agreement with the typical radio properties of known BL Lacs and FSRQs (i.e.\ the radio flux density distribution of the BCUs classified by us matches with that of the already classified BL Lacs and FSRQs) and (2) our method is more powerful than a simple analysis of the radio properties (i.e.\, there are many BCUs that can be classified as BL Lacs or FSRQs based on the ANN method, but would remain uncertain if we only looked at their radio flux density). 

\begin{figure}
\resizebox{\hsize}{!}{\includegraphics{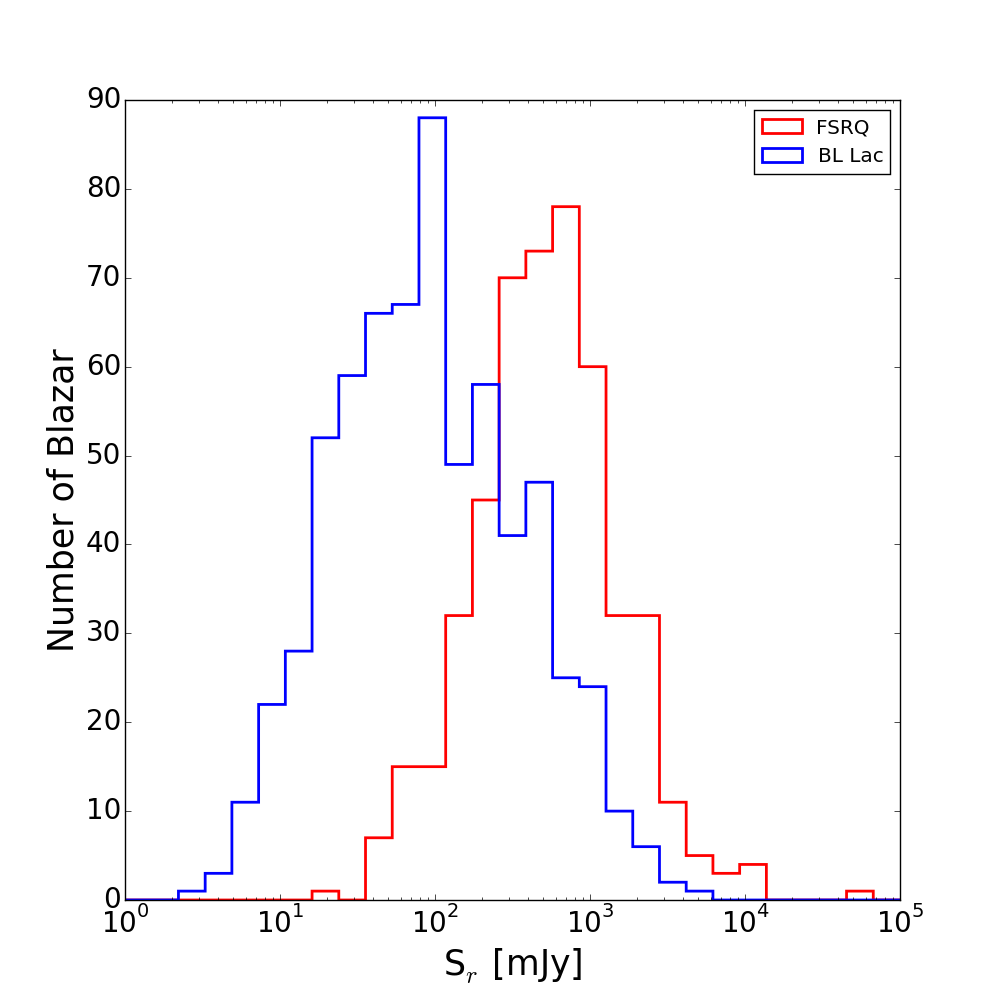}} 
\caption{$S_r$ distribution for the 3FGL sources classified as BL Lacs (blue histogram) and FSRQ (red histogram).}
\label{FigRadio3FGLHisto}
\end{figure}

Since blazars are, nearly by definition, radio-loud sources, radio flux densities for all of them can be readily obtained from large sky surveys. In particular, the 3LAC reports the radio flux density at 1.4 GHz from the NRAO VLA Sky Survey \citep[NVSS,][]{Condon1998} or at 0.8 GHz from the Sydney University Molonglo Sky Survey \citep[SUMSS,][]{Bock1999} for blazars located at Dec.~$>-40^\circ$ or $<-40^\circ$, respectively. In very few cases (only 20 in the entire clean 3LAC), radio flux densities are obtained at 20 GHz from the Australia Telescope Compact Array. In any case, blazars are flat-spectrum sources, and the error associated to assuming that $\alpha=0$ (i.e.\ treating all data as if they were taken at the same frequency) is not expected to be large. Hereafter, we indicate with $S_r$ the radio flux density, regardless of the source catalog.

\begin{figure*}
\resizebox{\hsize}{!}
{\includegraphics{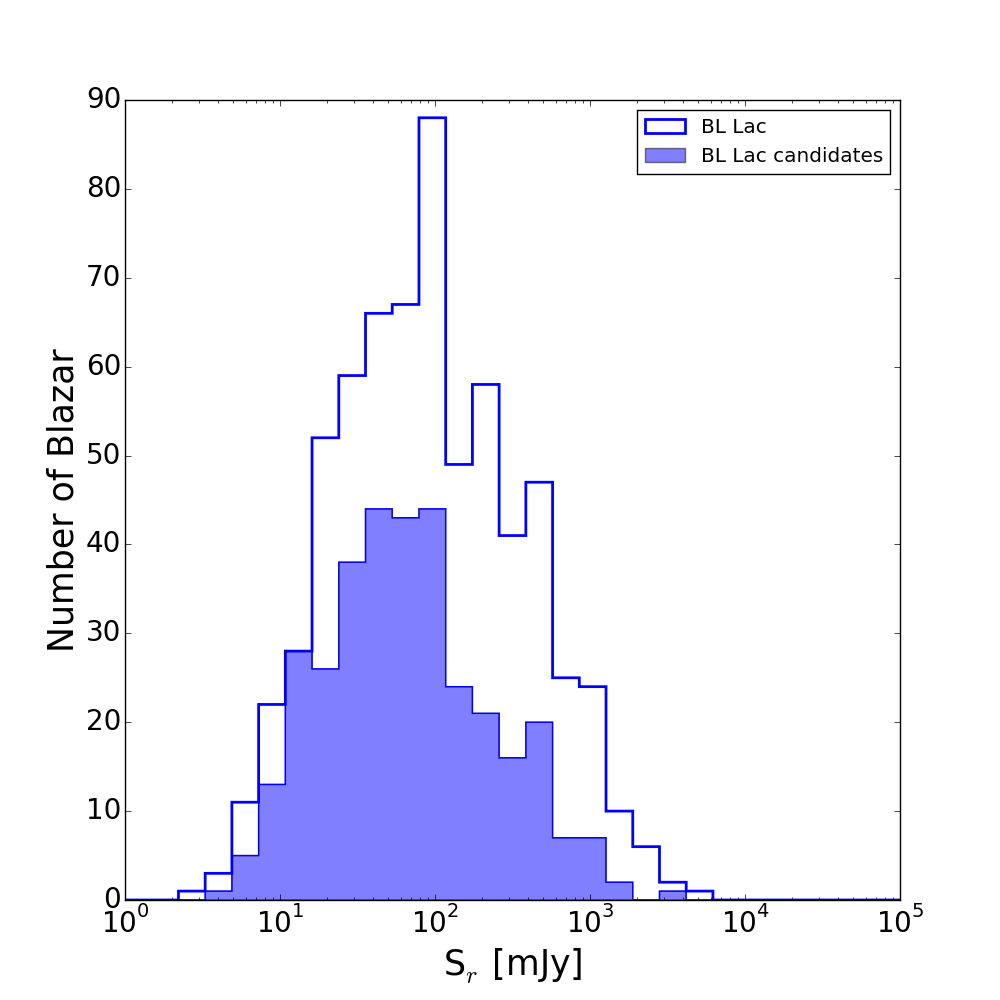}\includegraphics{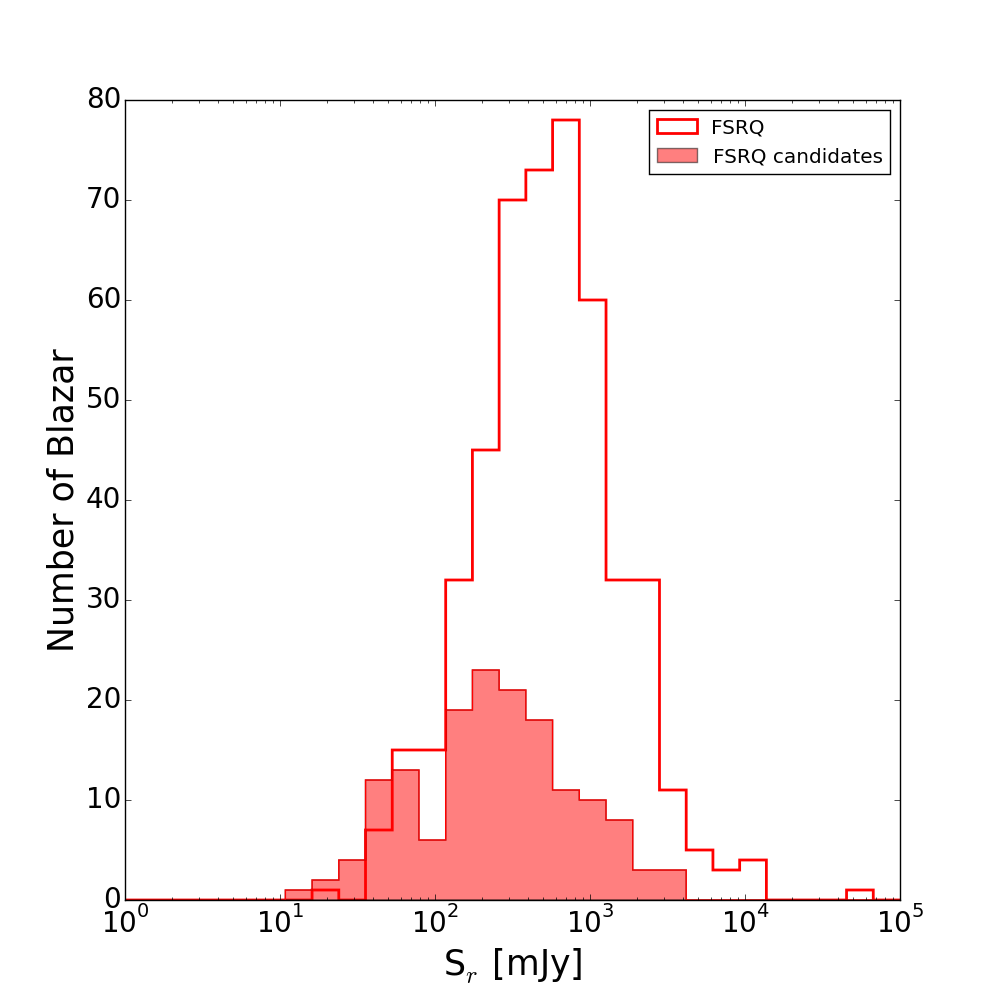}} 
\caption{$S_r$ distribution for the blazars classified through the \bflapann\ method (shaded histograms) in comparison to the previously classified blazars. Left: BL Lacs; right: FSRQs.}
\label{FigRadioAnnHisto}
\end{figure*}

In Fig.~\ref{FigRadio3FGLHisto}, we show the distribution of $S_r$ over the entire range of BCU flux densities, dividing between BL Lacs (blue histogram) and FSRQs (red histogram). The overall distribution is clearly bimodal, with BL Lacs peaking at lower flux density than FSRQs. Based on these distributions, we define two \emph{clean areas} where the density of sources of one class is predominant with respect to the other and  where it is possible to separate BL Lacs and FSRQs with a $90\%$ degree of purity.  These areas are defined by the thresholds $S<140$ mJy (90\% probability of being a BL Lac) and $S>2300$ mJy (90\% probability of being a FSRQ). We further note that there is only one FSRQs with $S<35$ mJy (while there are 170 BL Lacs in the same interval), corresponding to a \emph{superclean area} with 99.5\% probability of being a BL Lac. On the other hand, the overlap in the high flux density region is much larger and the radio flux density is not as reliable a predictor when it comes to identifying FSRQs.

In Fig.~\ref{FigRadioAnnHisto}, we compare the $S_r$ distribution for the sources classified through the \bflapann\ method (\bann\ and \fann, shown by shaded histograms) with that of the sources already classified in the 3FGL (\bfgl\ and \ffgl, shown by the empty histograms). In the left panel, we show the BL Lacs, in the right panel the FSRQs. It is readily seen that the radio flux density distributions  are in good agreement, which confirms the validity of our classification. In general, the \bflapann\ classified sources tend to lie on the fainter end of the distribution; that is not a surprise, since the brightest sources are more likely to have been selected for optical spectroscopy in past projects and therefore were not part of the starting BCU list.

In Figs.~\ref{FigRadioANNPlane}, we plot the ANN likelihood of a BCU being a BL Lac against $S_r$, divided in blocks according to the classification as a BL Lac or a FSRQ based on the ANN method and on the radio flux density. The blocks along the diagonal are those where the two methods agree, and they contain over 50\% of the total population of BCUs (295/573). Then, there is a large fraction ($190/573$, i.e.\ $\sim33\%$) of BCUs for which the ANN method provides a classification, while {\bf that based on} $S_r$ remains uncertain; these are the top and bottom blocks of the central column. This highlights the power of the ANN method in comparison to the simple flux density: only $\sim6\%$ of the BCUs can be classified through $S_r$ while they would remain uncertain for ANN. Finally, there is a $\sim8\%$ of sources for which the two methods disagree (top right and bottom left squares). These are probably quite peculiar objects or spurious associations that deserve a dedicated analysis beyond the scope of this paper. We further note that the analysis based on radio flux density could be subject to outliers, and in particular sources in the bottom left corner could be dim FSRQs that are located at very large redshift.

{\bf \begin{figure}
\resizebox{\hsize}{!}{\includegraphics{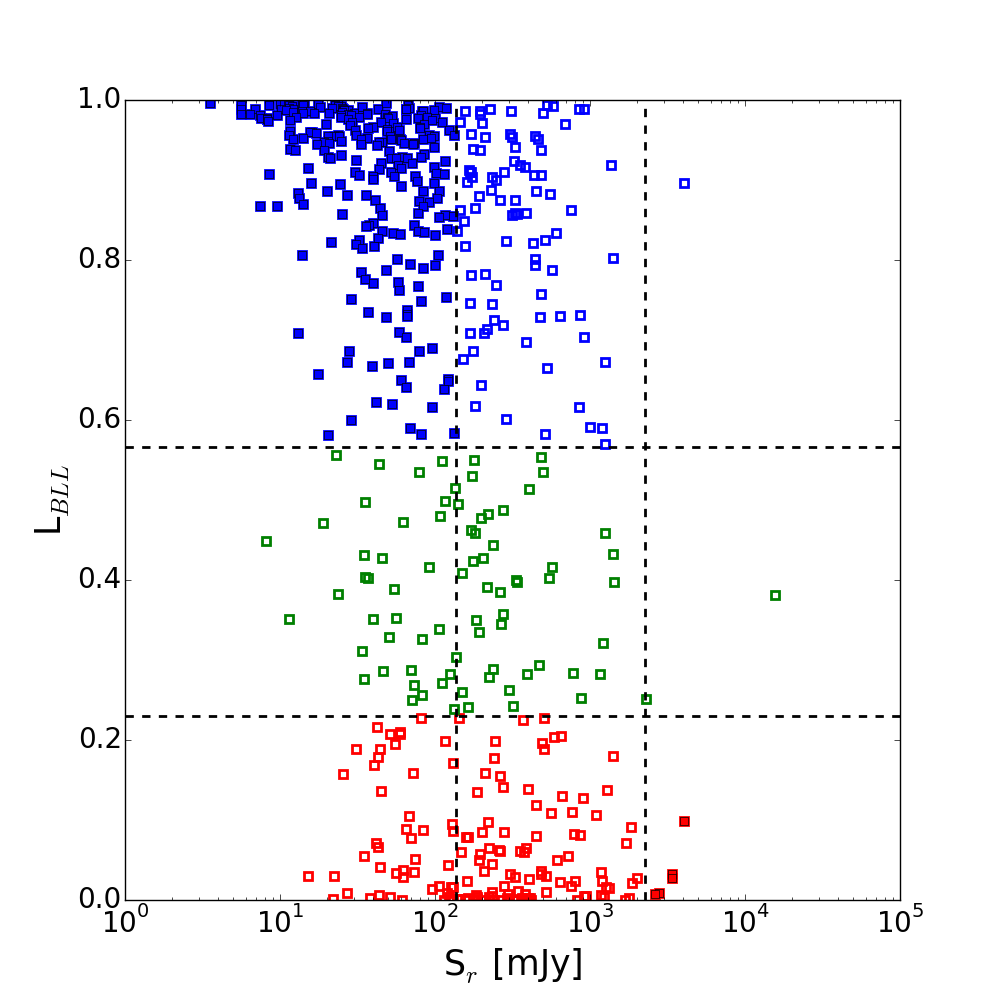}}
\caption{ ANN likelihood against $S_r$ in units of mJy. Colors indicate the classification proposed by the \bflapann\ method: blue for BL Lacs, red for FSRQs, and green for still uncertain objects. Filled symbols indicate the sources for which the radio flux density indicates a matching classification.}
\label{FigRadioANNPlane}
\end{figure}}

\section{Conclusion}
\label{sec:conclusions}

We developed an Artificial Neural Network technique based on B-FlaP information to assess the likelihood that a $\gamma$-ray source can belong to the BL Lac or FSRQ family, or more interestingly to the HSP blazar subclass, using only $\gamma$-ray data.
We tested the method on sources that are classified as BL Lacs or FSRQs in 3FGL, and focusing our attention on the HSP blazars, we found, and here we confirm,  that the method is very effective in the identification of blazars and offers an opportunity to provide a tentative HSP classification. 
This paper presents a full classification list of the 3FGL BCUs according to their ANN likelihood. BCUs can be divided in 342 BL Lac candidates (\bann), 154 FSRQ candidates (\fann) leaving only 77 without a clear classification. Among 573 BCUs we selected and ranked 53 very interesting HSP candidates to be observed through Very High Energy Telescope.
In order to validate the method we compared B-FlaP with the Variability Index and the  Power Law Index. In both the comparisons  B-FlaP showed full consistency, and  in some cases the efficiency of B-FlaP is greater than what is obtainable by the other parameters. 
To further assess the reliability of the method we performed direct optical observations for a sample of BCUs with Galactic latitude $|b| > 10^{\rm o}$ and maximum $\gamma$-ray flux less than $6 \cdot 10^{-8}\, {\rm ph\, cm^{-2}\, s^{-1}}$ . In those cases where we were able to perform spectroscopic observations we found that the optical spectra were fully consistent with the expectations based on the ANN results. 
Even the results of benchmarking between the radio data and B-FlaP showed a consistency of assessment with the two approaches.

We conclude that, although B-FlaP cannot replace confirmed and rigorous spectroscopic techniques for blazar classification, it may be configured as an additional powerful approach for the preliminary and reliable identification of BCUs and in particular the HSP blazar subclass when detailed observational or multiwavelength data are not yet available.

\section{Acknowledgments}
Support for science analysis during the operations phase is gratefully acknowledged from the \emph{Fermi}-LAT collaboration for making the 3FGL results available in such a useful form, the Institute of Space Astrophysics and Cosmic Physics of Milano -Italy (IASF INAF), and the Radioastronomy Institute INAF in Bologna Italy. Part of this work is based on observations collected at Copernico (or/and Schmidt) telescope(s) (Asiago, Italy) of the INAF - Osservatorio Astronomico di Padova. DS acknowledges support through EXTraS, funded from the European Commission Seventh Framework Programme (FP7/2007-2013) under grant agreement n. 607452.\\
The authors would like to thank the anonymous referee for discussion and suggestions leading to the improvement of this work.

\begin{figure*}
\begin{center}
\includegraphics[width=0.46\textwidth]{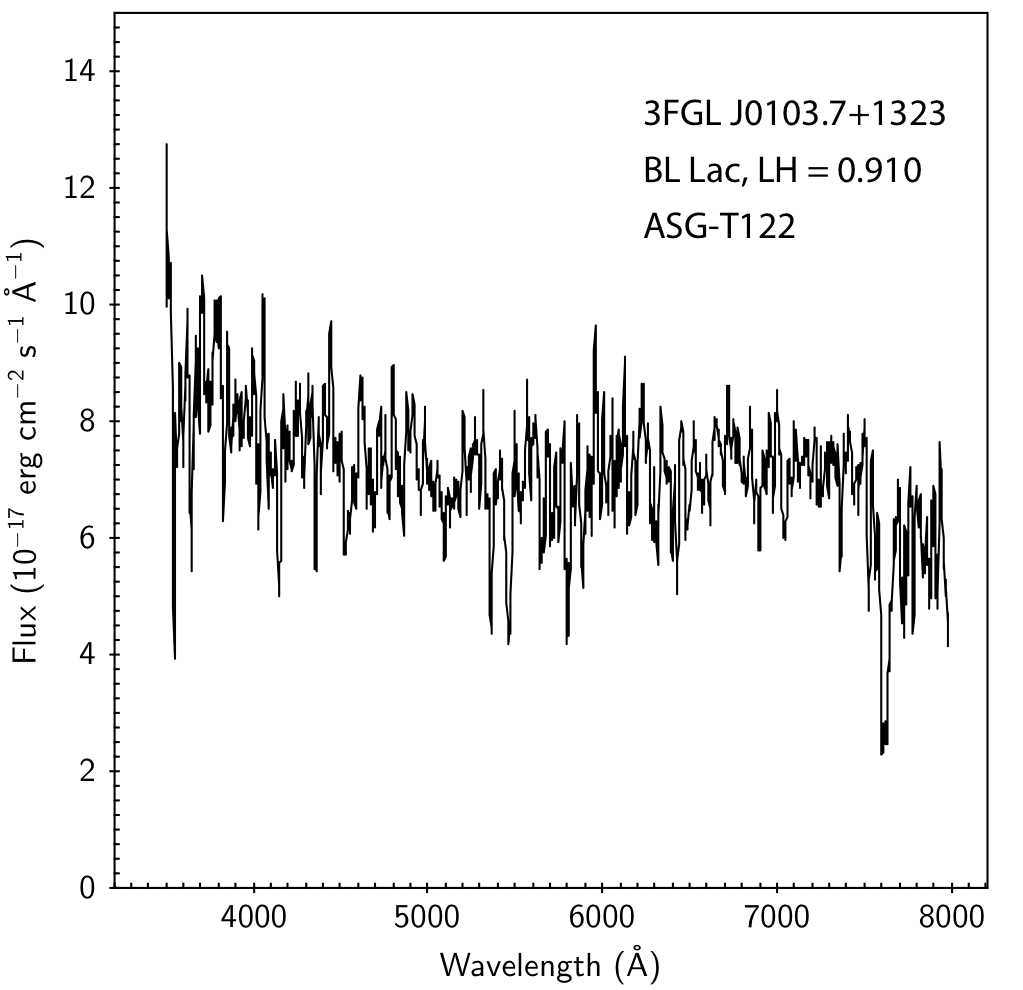} 
\includegraphics[width=0.46\textwidth]{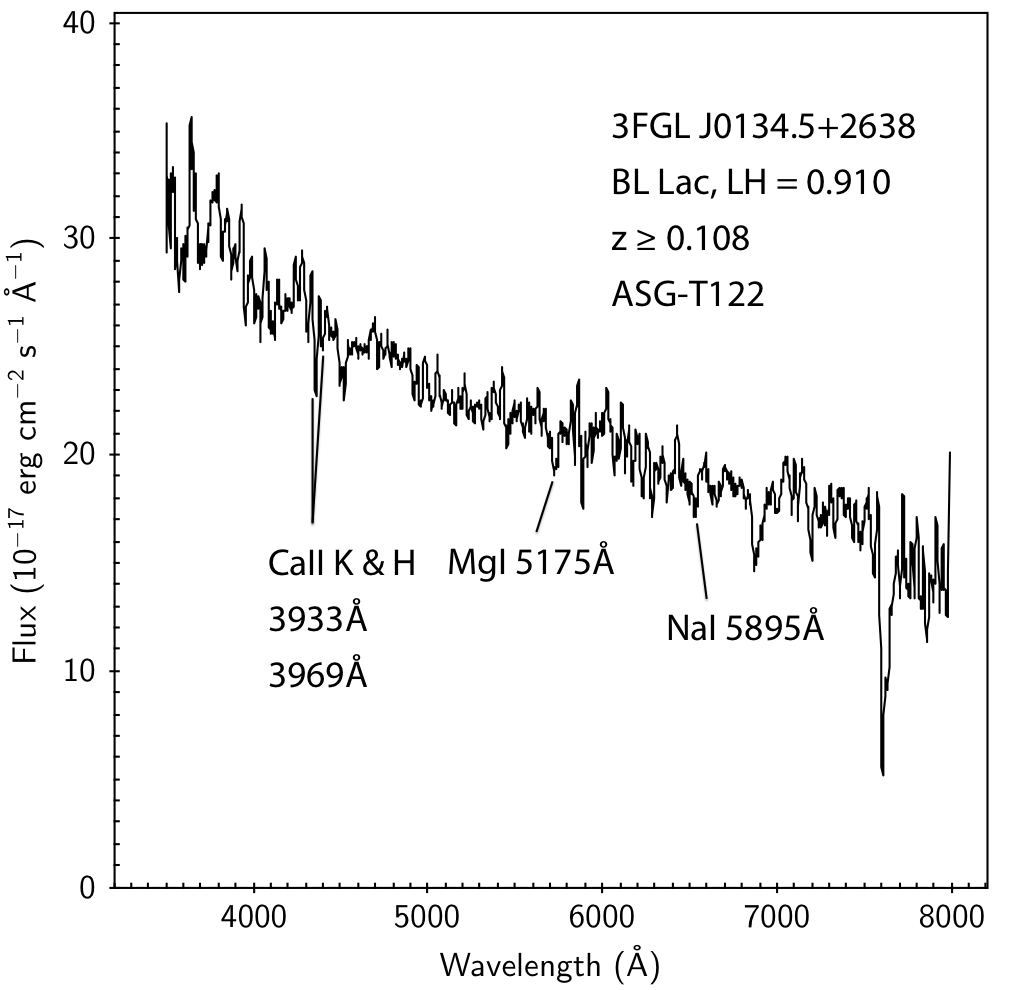}
\includegraphics[width=0.46\textwidth]{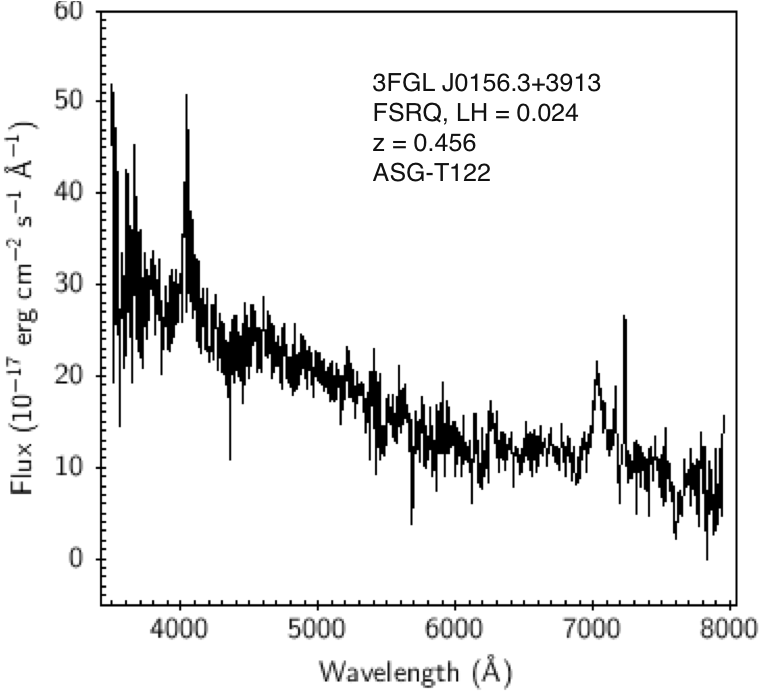} 
\includegraphics[width=0.46\textwidth]{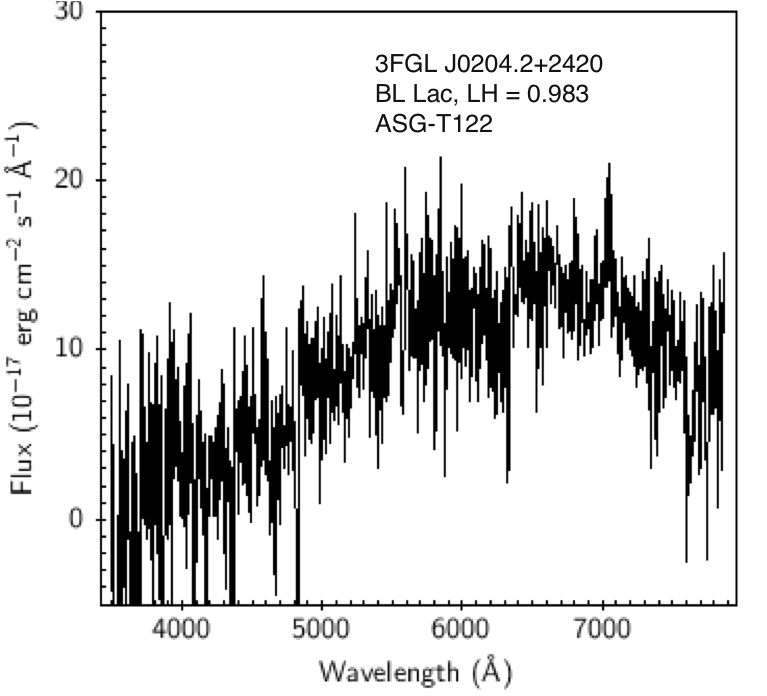}
\includegraphics[width=0.46\textwidth]{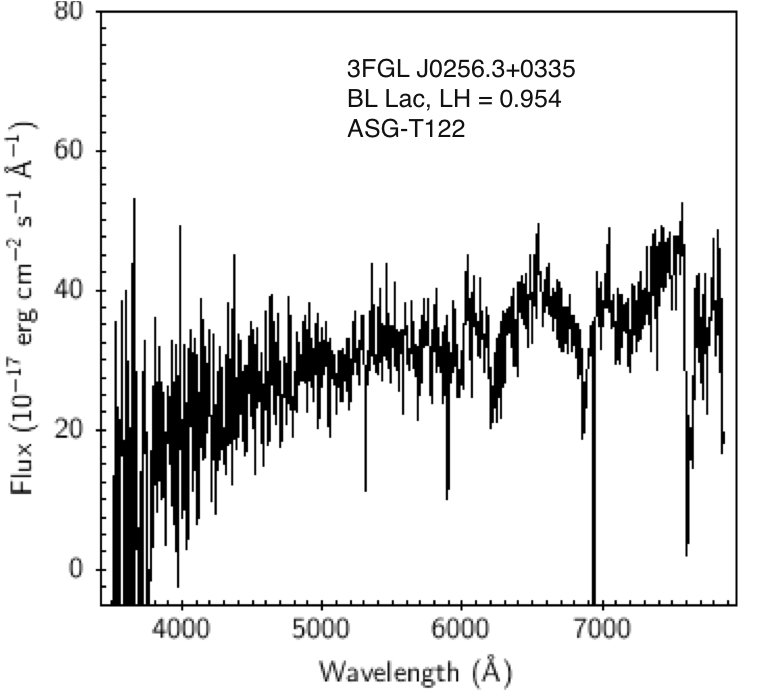}
\includegraphics[width=0.46\textwidth]{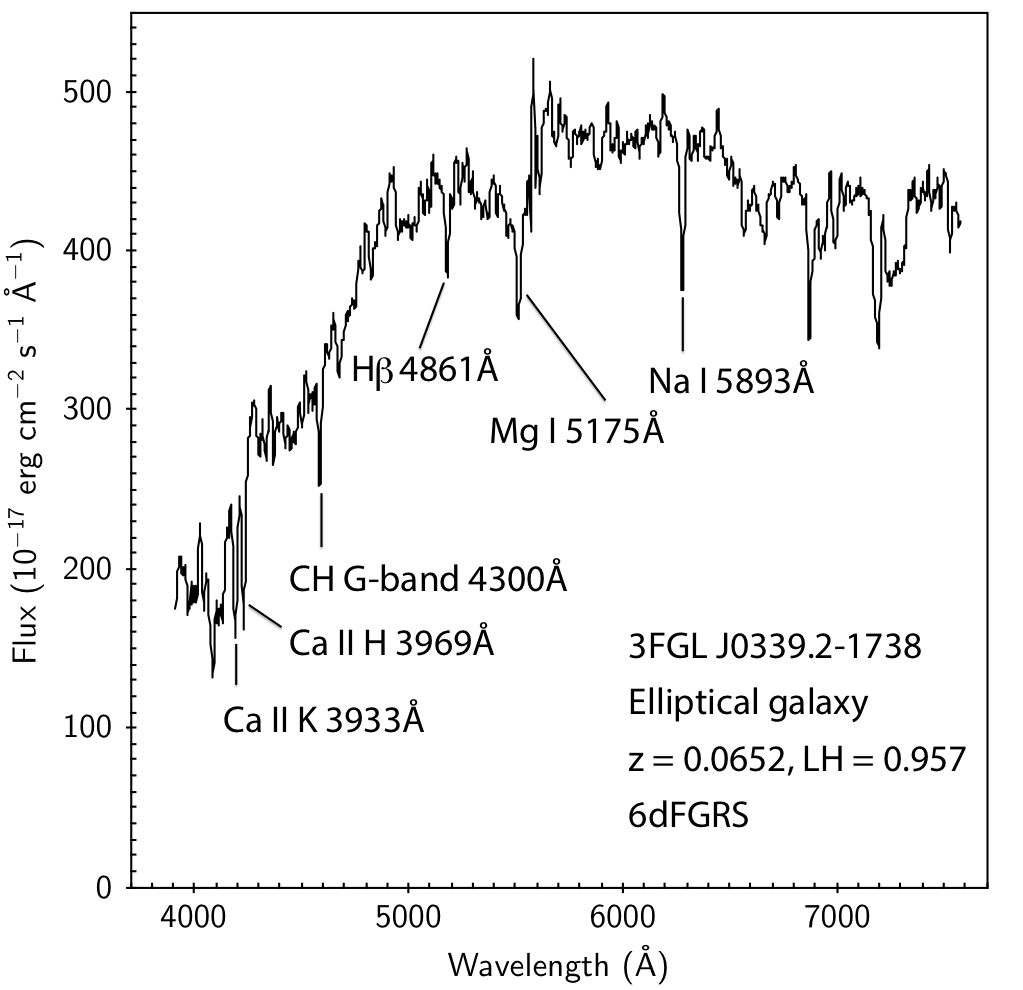}
\end{center}
{\bf Fig. \ref{FigOptSpec}} -- continued.
\end{figure*}

\begin{figure*}
\begin{center}
\includegraphics[width=0.46\textwidth]{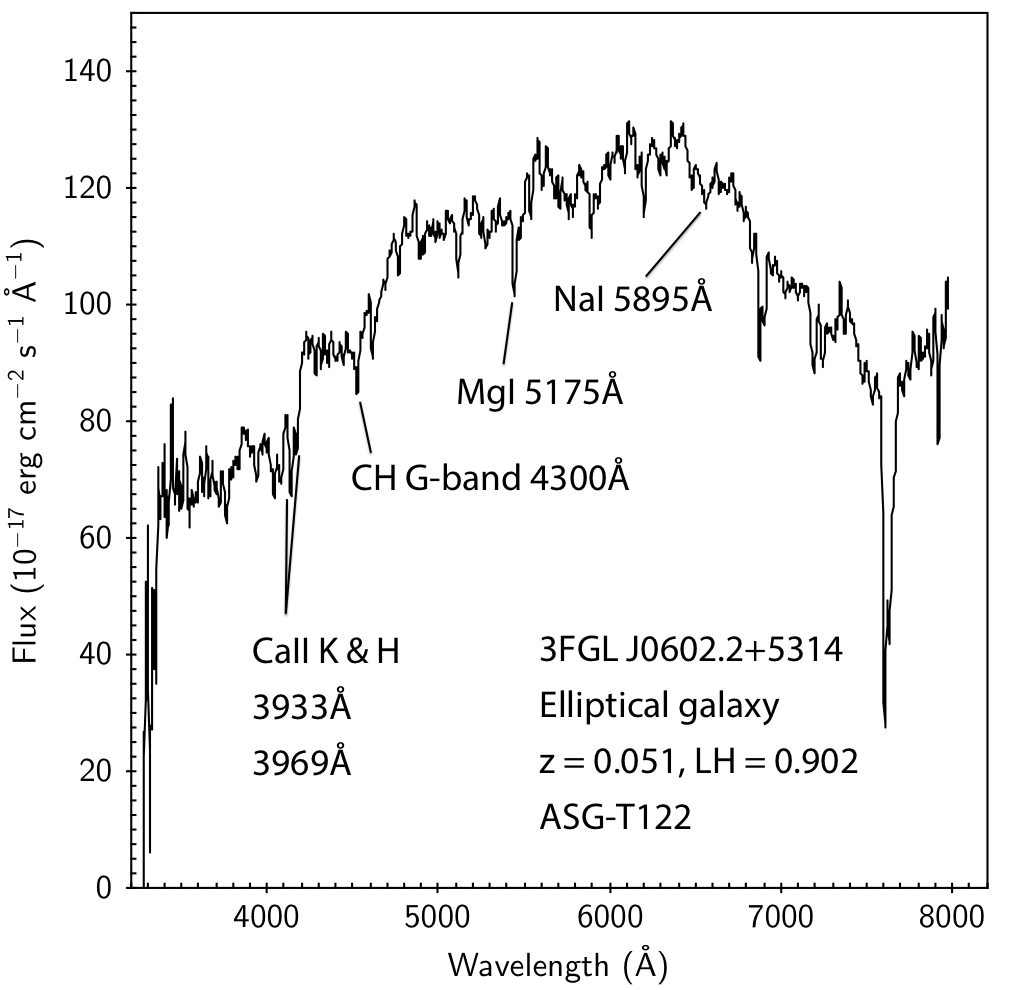}
\includegraphics[width=0.46\textwidth]{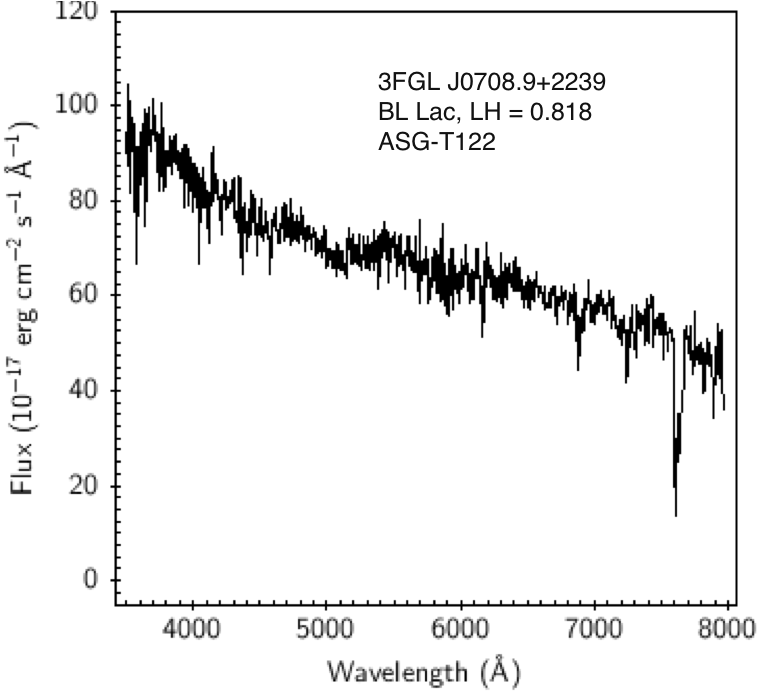}
\includegraphics[width=0.46\textwidth]{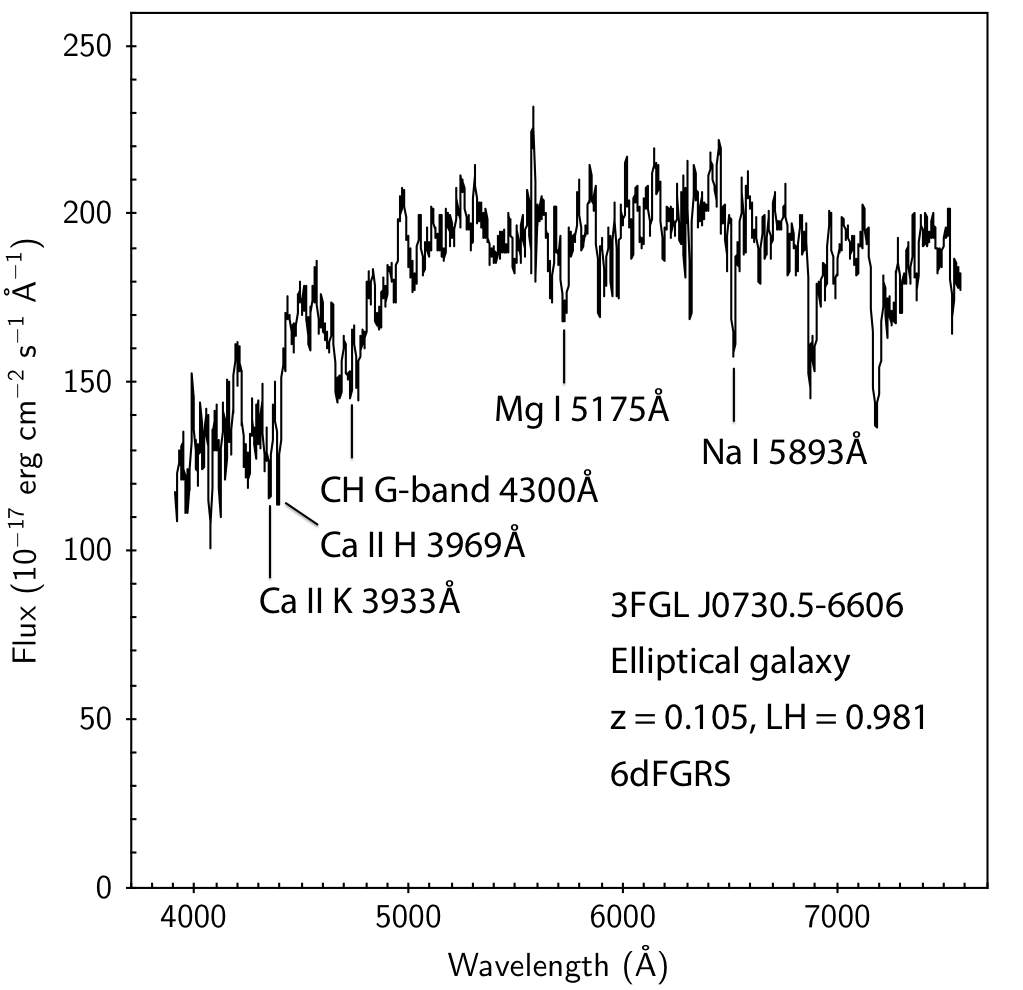}
\includegraphics[width=0.46\textwidth]{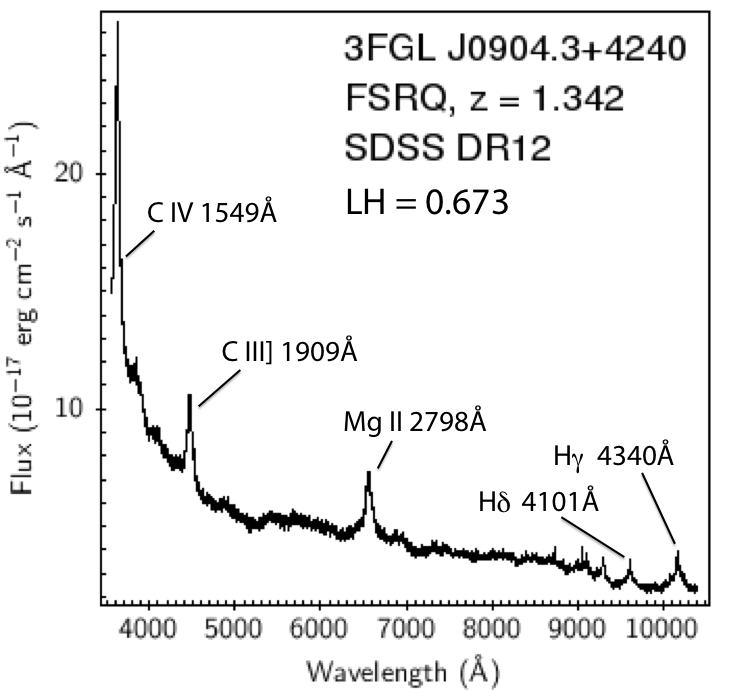}
\includegraphics[width=0.46\textwidth]{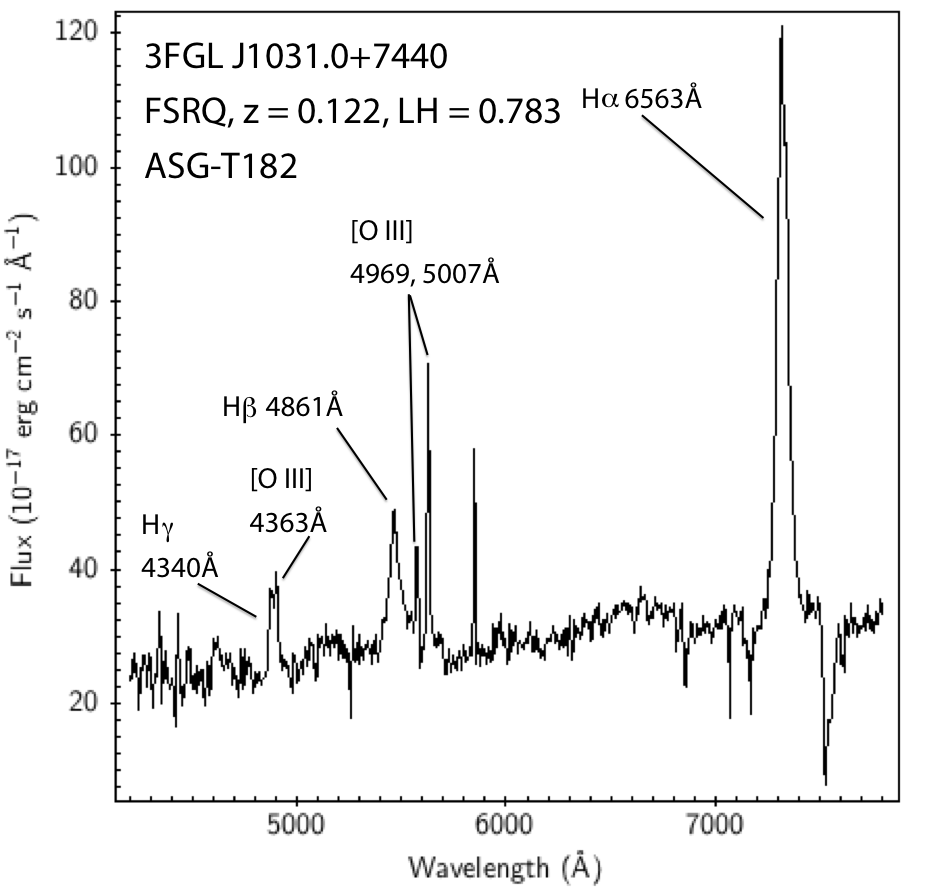}
\includegraphics[width=0.46\textwidth]{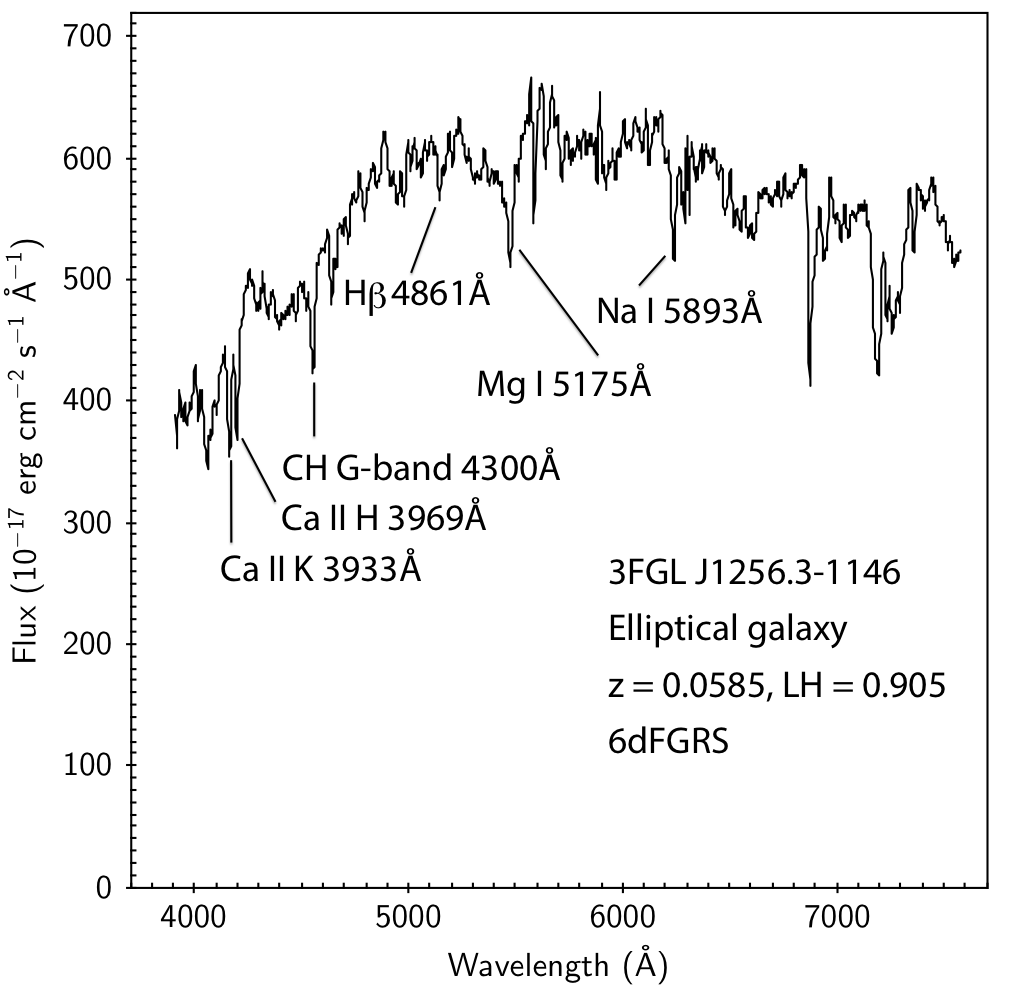}
\end{center}
{\bf Fig. \ref{FigOptSpec}} -- continued.
\end{figure*}

\begin{figure*}
\begin{center}
\includegraphics[width=0.46\textwidth]{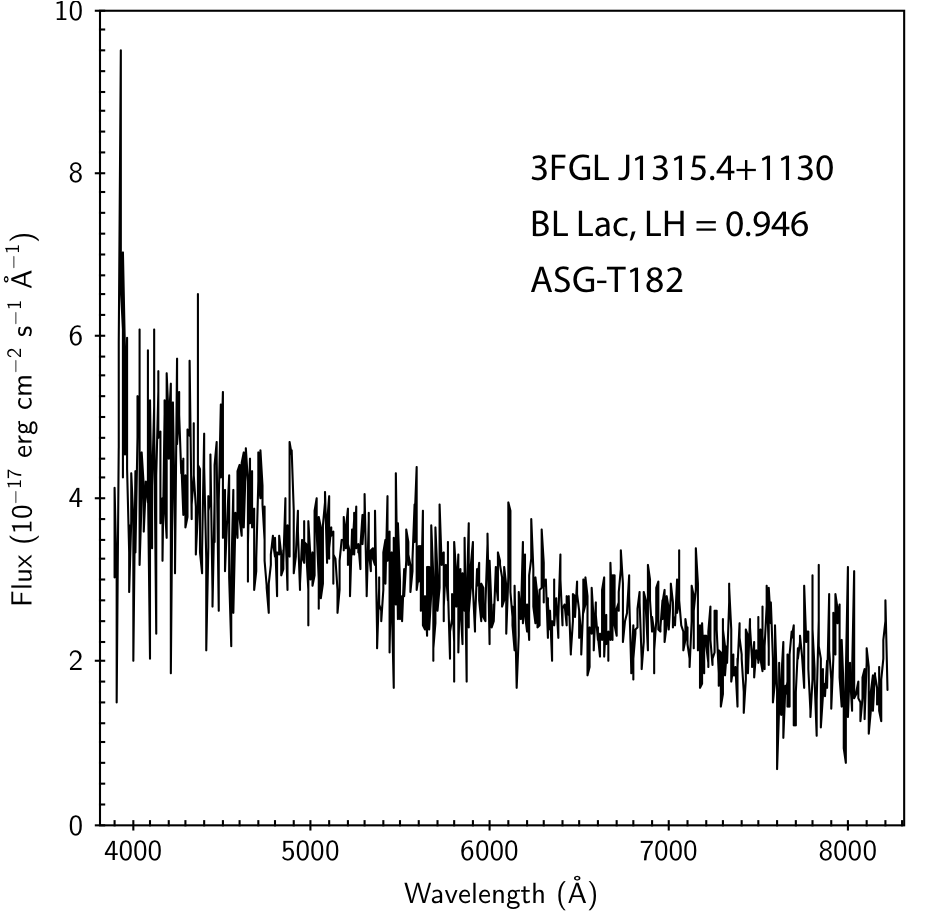}
\includegraphics[width=0.46\textwidth]{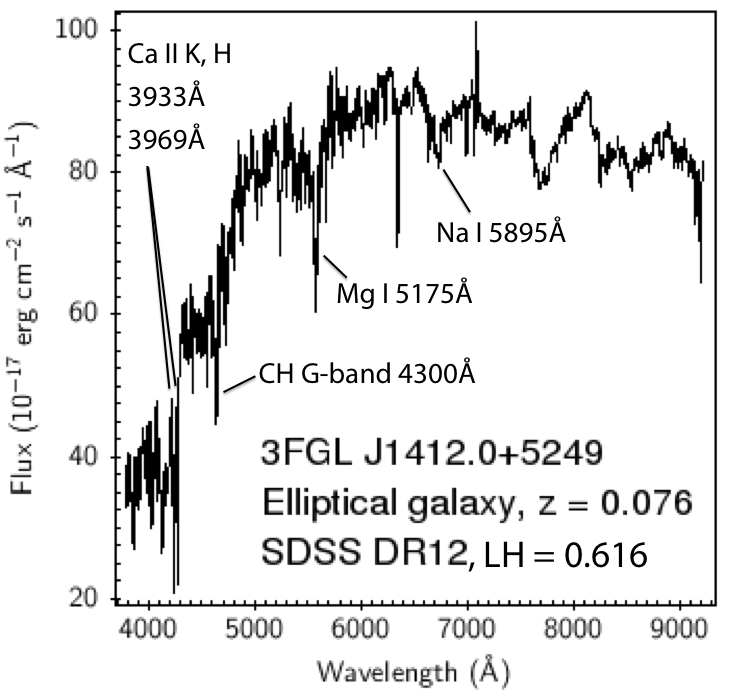}
\includegraphics[width=0.46\textwidth]{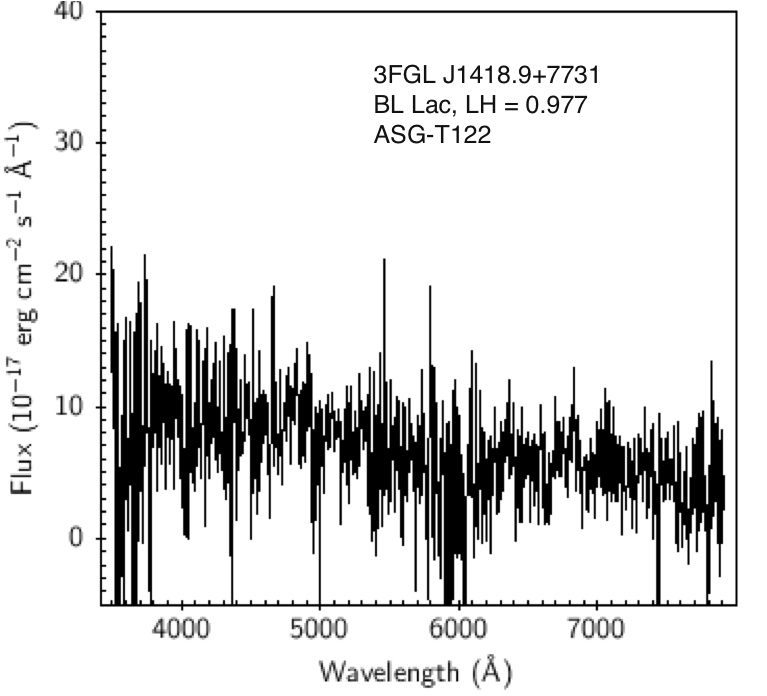}
\includegraphics[width=0.46\textwidth]{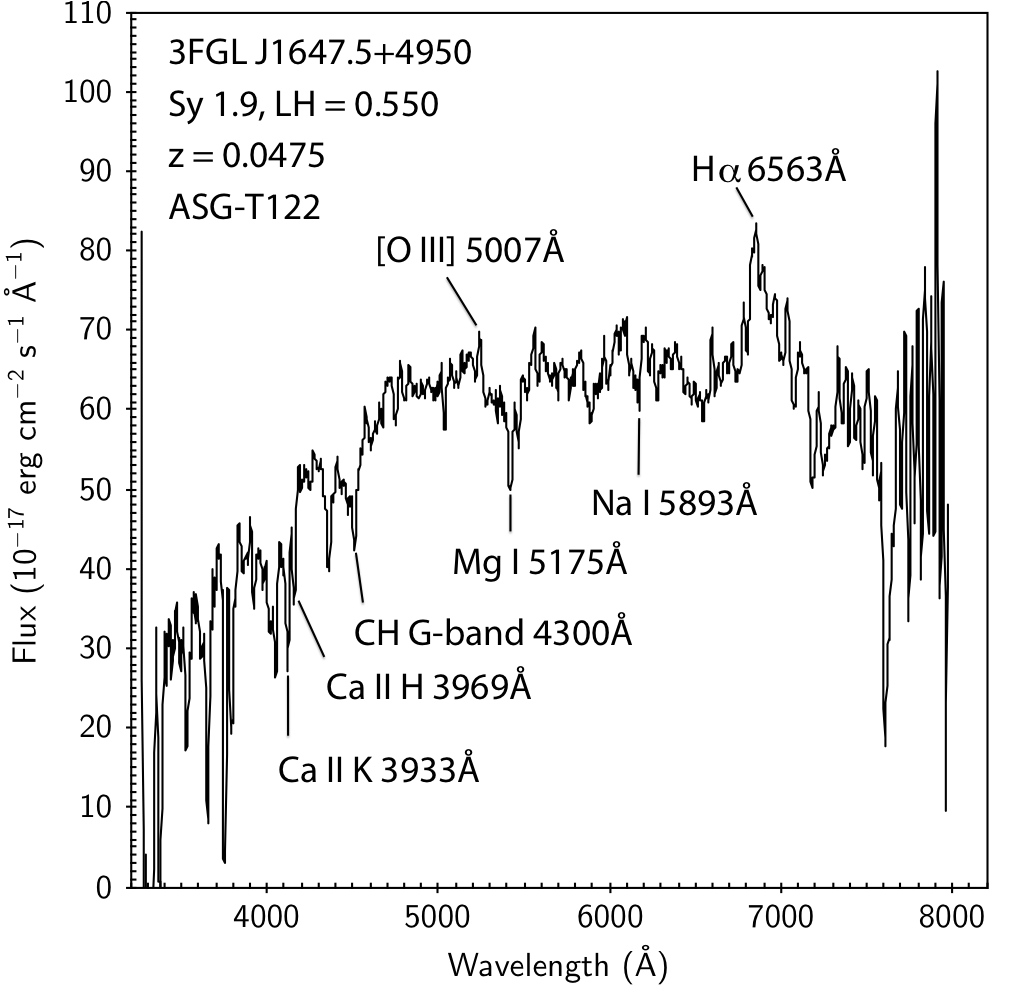}
\includegraphics[width=0.46\textwidth]{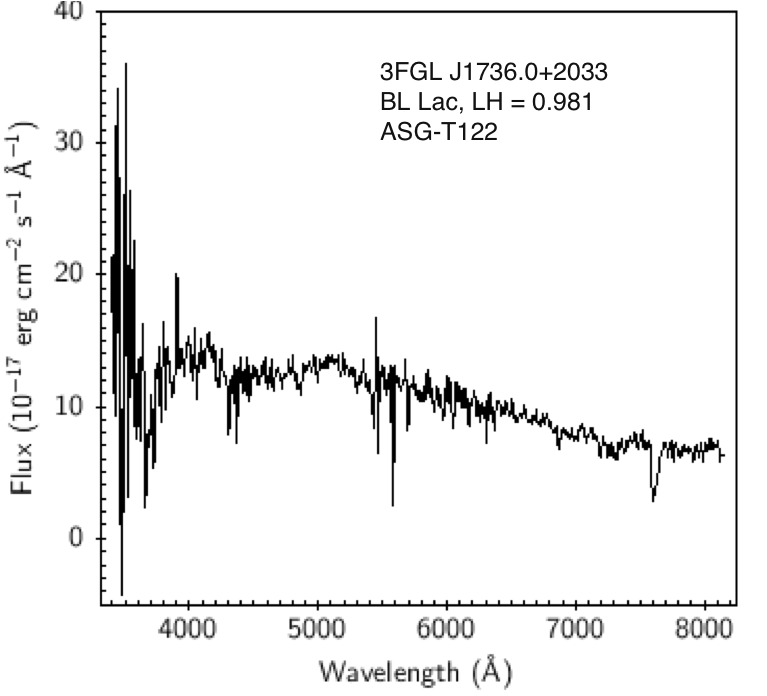}
\includegraphics[width=0.46\textwidth]{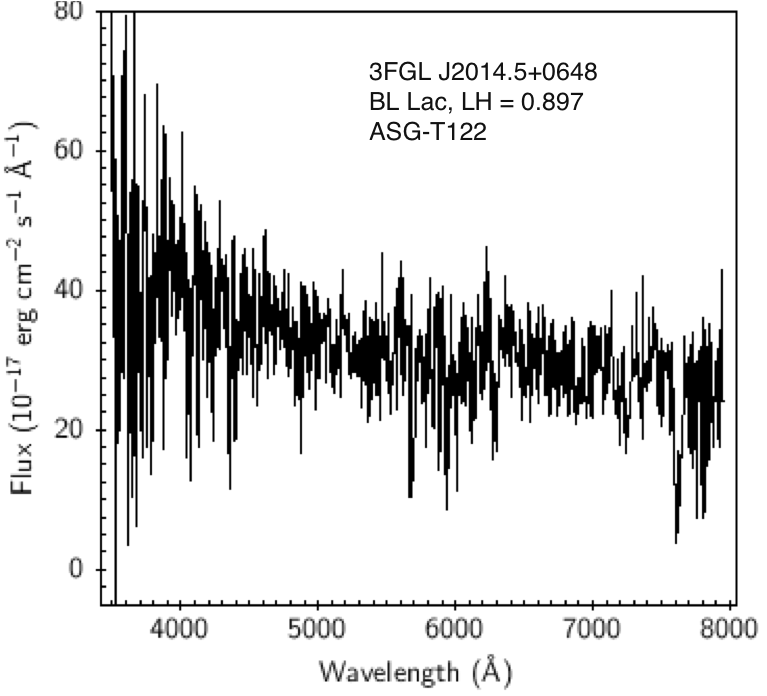}
\end{center}
{\bf Fig. \ref{FigOptSpec}} -- continued.
\end{figure*}

\bsp	
\label{lastpage}

\begin{thebibliography}{999} 

\bibitem[Abdo et al. (2010)]{Benoit01} Abdo A. et al., 2010, ApJ 710, 1271 

\bibitem[Abdo et  al. (2011)]{Abdo01} Abdo A. et al., 2011, ApJ, 716, 30 


\bibitem[Abdo et  al.(2010)]{Abdo02} Abdo  A. A.,
et al. , 2011, ApJ, 716, 30

\bibitem[Acero et  al.(2015)]{3FGL} Acero F.
et al. , 2015, ApJS, 218, 23

\bibitem[Acero et al.(2015)]{ace15} Acero F. et al., 2015, ApJ, 218, 23 

\bibitem[Ackermann et al.(2011a)]{Ackermann2011} Ackermann, M., 
et al.\ 2011a, ApJ, 741, 30 

\bibitem[Ackermann et al.(2011b)]{2LAC} Ackermann M., 
et al., 2011b, ApJ, 743, 171

\bibitem[Ackermann et al.(2015)]{2LACe} Ackermann M.
et al., 2015, ApJ, 806, 144


\bibitem[Ackermann et al. (2011)]{2FGL} Ackermann M. et al., 2011, ApJ 743, 171 

\bibitem[Ackermann et al. (2012)]{ack12} Ackermann M. et al., 2012, ApJ, 753, 83 

\bibitem[Ackermann et al.(2015)]{Ackermann2015} Ackermann M.
et al.,2015, ApJ, 810, 14 

\bibitem[Ahn et al. (2014)]{Ahn}Ahn C.P.
et al. , 2014, ApJ, 211,17

\bibitem[Ajello et al.(2014)]{Ajello} Ajello M.
et al., 2014, ApJ, 780,73

\bibitem[Ajello et al.(2014)]{Ajello} Ajello M. et al., 2014, ApJ, 780,73 

\bibitem[Ajello et al.(2016)]{ajello} Ajello M. 
et al., 2016, 
AAS Meeting, 227, id.403.04

\bibitem[Alam et al. (2015)]{Alam15} Alam S. 
et al., 2015, ApJS, 219, 12

\bibitem[Alvarez et al.(2016)]{alvarez}Alvarez Crespo N. 
et al., 2016,  The Astronomical Journal, Vol. 151, Issue 2, art. id. 32 

\bibitem[Atwood et al. (2009)]{FLAT} Atwood W. B.
et al. , 2009, ApJ, 697, 1071


\bibitem[Bishop (1995)]{bis95} Bishop C.~M., 1995, 

\bibitem[Bock et al.(1999)]{Bock1999} Bock D.~C. et al.,
1999, AJ, 117, 1578 

\bibitem[Condon et al.(1998)]{Condon1998} Condon, J.~J. 
et al.\ 1998, AJ, 115, 1693 

\bibitem[D'Ammando et al.(2015)]{D'Ammando2015} D'Ammando F. 
et al., 2015, MNRAS, 450, 3975 

\bibitem[De Naurois et al. (2015)]{deNaurois} De Naurois, M.   
et al., 2015, Comptes rendus - Physique, 16,  610

\bibitem[Doert et al. (2014)]{doe14} Doert M., and Errando M., 2014, ApJ, 782, 41

\bibitem[Doyle et al. (2005)]{Doyle05} Doyle M.~T.
et al., 2005, MNRAS, 361, 34


\bibitem[Ghisellini (2013)]{Ghisellini}Ghisellini G., 2013, 
EPJ Web of Conference, Vol. 61, id.05001.

\bibitem[Ghirlanda et al.(2010)]{Ghirlanda} Ghirlanda, G., et al., 2010, MNRAS, 413, 2, 852-862

\bibitem[Gish (1990)]{gis90} Gish H., 1990, Proceeding on Acoustic Speech and Signal Processing, pag. 1361

\bibitem[Hassan et al. (2013)]{has13} Hassan T. 
et al., 2013, MNRAS, 428, 220

\bibitem[Healey et al. (2007)]{Healey07} Healey S.~E.
et al, 2007, ApJS, 171, 61

\bibitem[Horan et al.(2008)]{horan}Horan D., Wakeley S., 2008, AAS, HEAD meeting 10, id.41.06


\bibitem[Jones et al. (2004)]{Jones04} Jones D.~H. 
et al. 2004, MNRAS, 355, 747

\bibitem[Jones et al. (2005)]{Jones05} Jones D.~H. 
et al., 2005, PASA, 22, 277

\bibitem[Jones et al. (2009)]{Jones09} Jones D.~H.
et al., 2009, MNRAS, 399, 683

\bibitem[Kolmogorov (1933)]{KS} Kolmogorov A. 
1933, G. Ist. Ital. Attuariale 4, p. 83–91

\bibitem[Lee et al. (2012)]{lee12} Lee K.~J.
et al.,2012, MNRAS, 424, 2832

\bibitem[Lott et al. (2012)]{Benoit02} Lott B. et al., 2012, A\&A, 544, A6 

\bibitem[Massaro et al.(2012)]{Massaro_classify} Massaro, F. 
et al., 2012 , ApJ 752, 61

\bibitem[Massaro et al.(2015)]{Massaro_review} Massaro, F., Thompson, D. J., Ferrara, E. C., 2015 , Astr. Ap. Review, in press

\bibitem[Mattox et al.(1996)]{Mattox} Mattox J.R.
et al., 1996 , ApJ, 461, 39


\bibitem[Meyer et al. (2011)]{Meyer11} Meyer E. et al., 2011, ApJ, 740, 98 

\bibitem[Mirabal et al. (2012)]{mir12} Mirabal N.
et al., 2012, MNRAS, 424, 64

\bibitem[Nolan et al.(2012)]{2FGL} Nolan P. L.
et al., 2012, ApJ Supplement Series , 199, 31  

\bibitem[Paggi et al. (2011)]{Paggi01} Paggi A. et al., 2011, ApJ, 736, 128 

\bibitem[Reed (1993)]{ree93} Reed R., 1993, IEEE Trans. Neural Networks, 4, 740

\bibitem[Richard et al. (1991)]{ric91} Richard M.~P., and Lippman R.~P., 1991, Neural Computation, 3, 461

\bibitem[Rojas (1996)]{roj96} Rojas R., 1996, Neural Networks: A Systematic Introduction, Springer-Verlag

\bibitem[Ruan et al.(2012)]{VAR}Ruan J.
et al., 2012, ApJ, 760, 51

\bibitem[Santos et al.(2002)]{Santos02} Santos J.~F.~C.~Jr. 
et al., 2002, IAUS, 207, 727

\bibitem[Saz Parkinson et al. (2016)]{park16} Saz Parkinson P.~M. 
et al., 2016, ApJ, 820, 8

\bibitem[Urry \& Padovani(1995)]{padovani} C. M. Urry and P. Padovani, Publ. Astron. Soc. Pacific 107, 803 (1995)






















\end{thebibliography}
\end{document}